\documentclass{elsart}

\usepackage[square,comma]{natbib}%bibliography

\usepackage{latexsym}
\usepackage{amssymb}
\usepackage{amsmath}%
\usepackage{graphicx}%

\makeatletter
\renewcommand{\theequation}{\thesection.\arabic{equation}}
\@addtoreset{equation}{section}
\makeatother

\newcommand{\la}{\langle}
\newcommand{\ra}{\rangle}

\newcommand{\hphi}{{\hat \phi}}
\newcommand{\hpi}{{\hat \pi}}

\newcommand{\tphi}{{\tilde \phi}}

\newcommand \bp{{\mathbf p}}
\newcommand \bk{{\mathbf k}}

\newcommand \bq{{\mathbf q}}

\newcommand \bv{{\mathbf v}}

\newcommand{\caln}{{\cal N}}
\newcommand{\calg}{{\cal G}}

\newcommand{\DPi}{{\Delta \Pi}}

 \def\be{\beta}  
   \def\ep{\epsilon}
\def\dg{\dagger}
\def\hphi{\hat{\phi}}
\def\hpi{\hat{\pi}}

\def\tphi{\tilde{\phi}}
 
\def\v#1{\mathbf{#1}}			%
\def\r{\v{r}} 				% def. of vector "r"
\def\p{\v{p}} 				% def. of vector "p"
\def\k{\v{k}} 				% def. of vector "k"
\def\rt{(\v{r},t)}   			% (r,t)
\def\prt{(\v{p},\v{r},t)}		% (p,r,t)
\def\pd{\partial} 			
\def\vnab{\v{\nabla}}
\def\w{\omega} 				% def. of w as omega
\def\wp{\omega_{\v{p}}}		% ã?Ê?ã?Ê?_{p}
\def\wa{\omega_{1,\p}}	% ã?Ê?ã?Ê?_{1,p}
\def\wb{\omega_{2,\p}}	% ã?Ê?ã?Ê?_{2,p}
\def\phii{\phi_{i,c}}		% ã?Ê?ã?Ê?_{i,c}
\def\tphii{\tilde{\phi}_i}	% ~ã?Ê?ã?Ê?_i
\def\la{\langle}
\def\ra{\rangle}
\def\fla{\langle \tilde{\phi}^2_1 \rangle}	%<~ã?Ê?ã?Ê?1^2>
\def\flb{\langle \tilde{\phi}^2_2 \rangle}	%<~ã?Ê?ã?Ê?2^2>
\def\foi{f_{\rm eq.} (\omega_{i,\p}) } % f1,eq(p)
\def\foj{f_{\rm eq.} (\omega_{j,\p}) } % f2,eq(p)
\def\foa{f_{\rm eq.} (\omega_{1,\p}) } % f1,eq(p)
\def\fob{f_{\rm eq.} (\omega_{2,\p}) } % f2,eq(p)
\def\calf{\mathcal{F}}
\def\calk{\mathcal{K}}

\def\calj{\mathcal{J}}

\def\tr{\text{tr}}

\def\i{\mbox{i}} \def\e{\mbox{e}}

\def\be{\begin{eqnarray}}
\def\ee{\end{eqnarray}}

\begin{document}
\begin{frontmatter}


\rightline{UT-Komaba/08-21}
\rightline{KEK-TH-1291}

\title{Quantized meson fields in and out of equilibrium. II: Chiral condensate and collective meson excitations}
\author{M. Matsuo\ead{matsuo@post.kek.jp}}
\address{Institute of Physics, University of Tokyo,
Komaba, Tokyo 153-8902, Japan and 
Institute of Particle and Nuclear Studies, High Energy Accelerator Research Organization,
Tsukuba, Ibaraki 305-0801, Japan\thanksref{mm}}
\author{T. Matsui\ead{tmatsui@hep1.c.u-tokyo.ac.jp}}
\address{Institute of Physics, University of Tokyo \\
Komaba, Tokyo 153-8902, Japan}
\thanks[mm]{present address}

\begin{abstract}
We develop a quantum kinetic theory of the chiral condensate and meson quasi-particle excitations using the $O(N)$ linear sigma model which describe the chiral phase transition both in and out of equilibrium in a unified way. 
A mean field approximation is formulated in the presence of mesonic quasi-particle excitations which are described by generalized Wigner functions. It is shown that in equilibrium our kinetic equations reduce to the gap equations which determine the equilibrium condensate amplitude and the effective masses of the quasi-particle excitations, while linearization of transport equations, near such equilibrium, determine the dispersion relations of the collective mesonic excitations at finite temperatures. 
Although all mass parameters for the meson excitations become at finite temperature, apparently violating the Goldstone theorem, the missing Nambu-Goldstone modes are retrieved in the collective excitations of the system as three degenerate phonon-like modes in the symmetry-broken phase.
We show that the temperature dependence of the pole masses of the 
collective pion excitations has non-analytic kink behavior at the threshold 
of the quasi-particle excitations in the presence of explicit
symmetry breaking interaction. 
\end{abstract}
\end{frontmatter}

% main text
\section{Introduction}
In our previous paper~\cite{MM08}, hereafter referred to as I, we have formulated a 
kinetic theory of self-interacting meson fields with an aim to describe the freezeout stage of the 
space-time evolution of matter in relativistic heavy-ion collisions. 
By using a single component real scalar field model we have obtained a set of coupled equations 
of motion for the meson condensate and the quasi-particle excitations. 
The meson condensate obeys the classical field equation with a modification due to the coupling 
to mesonic quasi-particle excitations, which is  expressed in terms of the Wigner functions defined 
by quantum statistical expectation values of the bilinear forms of the quantized field operators. 
The Wigner functions contain the off-diagonal components which can be eliminated by Bogoliubov 
transformation in uniform systems, but survive in general non-uniform systems. 
We also solved the kinetic theory in and near equilibrium and studied a gap equation and 
the dispersion relation of the collective excitations. 

It is the purpose of the present paper to apply this formalism to the $O(N)$ symmetric linear sigma 
model which possesses a continuous symmetry. 
For $N=4$, this model may be considered as an effective field theory of QCD with light quark flavors:
the chiral condensate is described in terms of the sigma meson condensate, which is defined by the 
expectation value of the first component of the $N$-component scalar quantum fields.  

The new aspect which arises due to the presence of continuous symmetry is the manifestation 
of the Nambu-Goldstone modes\cite{Nam60,Gol61} which are identified as pions. 
The chiral phase transition has been studied using various effective chiral models of QCD, 
such as the linear sigma model\cite{BG77} with flavor extension\cite{PW84}, 
the Nambu-Jona-Lasinio model for effective quark fields\cite{HK85}, and the
non-linear realization of Weinberg as a low energy effecive theory of QCD\cite{GL87}.
Here we adopt the $O(N)$ sigma model which has been originally used in the context of 
electro-weak phase transition\cite{DJ74}.  
The model has been used by many authors to describe the chiral phase transition at finite temperature in the mean field (Hartree) approximations
\cite{BG77,RM98,A-C97,CH98,Pet99,NNO00}.  
We shall extend such descriptions to non-equiibrium situations with the formalism developed in I.   

In this paper we first derive a coupled kinetic equations for the chiral condensate and 
the mesonic quasi-particle excitation and show these equations reduce in equilibrium to the gap 
equations to determine the masses of the quasi-particle excitations in the Hartree approximation.
The solution of these gap equations exhibits a first order phase transition. 
This is a well-known generic problem of the mean field approximation which may be remedied by the inclusion of higher order fluctuations\cite{Chi00}.  
It has been known also, however, that such singular behavior of the chiral phase transition is 
smoothed out also by the introduction of the explicit symmetry breaking 
so that it generates the observed pion mass in the vacuum.  
In this work, we also study the effects of the explicit symmetry breaking in the present formalism. 

Another difficulty associated with the mean field approximation is that the pion quasi-particle 
excitation becomes massive apparently violating the Goldstone theorem \cite{Gol61}.  
This is another generic problem of the Hartree approximation. 
It has been known that the missing Nambu-Goldstone modes arise as acoustic collective modes of the excitations 
\cite{Okopinska1996,DSM96,NGP97,TVM00}. 
We demonstrate this by computing the dispersion relations of the collective excitations of the system near equilibrium
from the linearized kinetic equation with respect small deviations of the condensate amplitude and the 
quasi-particle distributions from their equilibrium values
These modes appears in the continuum of the quasi-particle excitations and thus may suffer a collisionless damping (Landau damping).  
We examine how these Nambu-Goldstoned modes are affected by the inclusion of the explicit chiral 
symmetry breaking.  
It will be shown that the Nambu-Goldstone mode becomes a massive excitation in 
the quasi-particle continuum which turn into un-damped pionic excitation in the high 
temperature symmetry recovered phase through a kink non-analytically. 
  
In the next section we extend the formalism developed in I to derive the coupled kinetic 
equations for $N$ component self-interacting fields with $O(N)$ symmetry. 
We will show that $N$ component meson condensate is described by $N$ non-linear 
Klein-Gordon equations which contain non-linear interaction terms 
given in terms of the $N$ classical mean fields and $N \times N$ component matrix 
of $N$ fluctuating fields.  
The quasi-particle degrees of freedom is expressed in terms of a $2 \times 2$ matrix of 
generalized Wigner functions which consists of the ordinary one-body density matrices 
in diagonal components and "anomalous" one-body density matrices in off-diagonal 
components, each component being $N \times N$ matrix of $N$ component fields. 
For the long wavelength excitations,  the equations of motion of the diagonal components 
of the generalized Wigner functions are shown to reduce to the Landau-Vlasov-type kinetic 
equations for the quasi-particle distribution function with a source/sink term expressed 
in terms of the off-diagonal components of the generalized Wigner functions. 
Kinetic equations similar to ours have been derived for dilute cold atomic gases
in \cite{ITG99}.  

In section 4, uniform static solutions of these coupled kinetic equations are shown to reproduce 
the well-known Hartree equilibrium solutions at finite temperature which exhibit the first order 
chiral phase transition.   All off-diagonal components of the generalized Wigner function vanish
in equilibrium and the $N$ diagonal components give thermal distributions of the $N$ 
kinds of quasi-particles, one being sigma-like and other $N-1$ pion-like modes.   

In section 5, we apply our coupled kinetic equations to compute the dispersion relations 
of the collective excitations near equilibrium. 
By linearizing the kinetic equations with respect to a small deviations from the static, uniform
equilibrium solution we derive a set of self-consistency conditions to determine the 
dispersion relation of the collective mode. 
It will be shown that the fluctuation of the diagonal components of $N \times N$ matrix
are all coupled and give dispersion relation of sigma-like collective mode while
pion-like collective modes are contained in the off-diagonal fluctuations of the 
Wigner functions.
The missing Nambu-Goldstone modes are retrieved in these pion-like collective
modes as acoustic modes by making use of the self-consistency conditions to
determine the condensate in equilibrium.
We examine how these collective modes are modified when the chiral symmetry is not exact
at the end of this section.   

A short summary of this work is given in the last section together with some open problems. 
The explicit expression of our kinetic equations in $O(N)$ model are given in the appendix.

\section{Kinetic equations for $O(N)$ linear sigma model}
In the previous paper\cite{MM08}, we discussed coupled kinetic equations for one-component real scalar field, 
which has no continuous symmetry. 
In order to apply for more realistic physical situation, we need to extend the model to $O(N)$ symmetric linear sigma model which 
possess continuous symmetry.
The new aspects arise due to the presence of continuous symmetry, namely the manifestation of the Nambu-Goldstone modes, 
can be seen in the $O(N)$ sigma model.

\subsection{Heisenberg equations of motion for $O(N)$ sigma model}
The Hamiltonian of $O(N)$ sigma model is given by
\begin{eqnarray}\label{Hamiltonian}
H=\int d \r \left[ \frac{1}{2} \sum_{i=1}^{N} \left( \hat{\pi_{i}}^2 + \left( \nabla \hat{\phi_i} \right)^2 +  m^2 \hat{\phi_i}^2 +h_i \hat{\phi_i} \right) + \frac{\lambda}{4!} \Bigl( \sum_{i=1}^{N} \hat{\phi_i}^2 \Bigr)^2  \right]
\end{eqnarray}
 where scalar fields $\hat{\phi}_i$ and their canonical conjugate momentum field $\hat{\pi}_i$ are quantized by the usual equal-time commutation relations:
\begin{eqnarray}
 \left[\hphi_i(\r,t),\hpi_j(\r',t) \right]&=& i \delta_{i,j}\delta (\r-\r') \\
 \left[\hphi_i(\r,t),\hphi_j(\r',t) \right] &=& \left[\hpi_i(\r,t),\hpi_j(\r',t) \right]=0.
\end{eqnarray}
In the absence of the external field $h_i$, $H$ is symmetric with respect to rotaions of $\phi_i$ and $\pi_i$ which form an $O(N)$ group.
This Hamiltonian may be considered as a meson sector of the effective Hamiltonian of QCD if we set $h_i=\epsilon \delta_{i,1}$ and $\epsilon$ is chosen to reproduce the pion mass.
$\phi_1$ corresponds to the sigma meson field and $(\phi_2,\phi_3,\phi_4)$ form isovector pion fields.

The Heisenberg equation of motion of the quantum fields $\hat{\phi}_i$ is given by
\begin{eqnarray}
\frac{\pd \hphi_i}{\pd t} = - \i \left[ \hphi_i, H \right] = \hpi_i(\r,t) 
\end{eqnarray}
and the equation of motion of the canonical conjugate field $\hpi(\r,t)_i$ becomes
\begin{eqnarray}
\frac{\pd \hpi_i}{\pd t } = - \i \left[ \hpi_i, H \right] = (\vnab^2 - m^2) \hphi_i \rt - \frac{\lambda}{3!}  \hphi_i \rt \sum_{j = 1}^N \hphi_j^2 \rt -h_i
\end{eqnarray}
After eliminating the momentum field $\hat{\pi}$ from these equations, we obtained a Klein-Gordon equation for the quantum scalar field $\hat{\phi}_i$:

\begin{equation}
\Box \hphi_i \rt + m^2 \hphi_i \rt = - \frac{\lambda}{3!} \hphi_i \rt \sum_{j = 1}^N \hphi_j^2 \rt -h_i  \label{eq:ON-KG}
\end{equation}

\subsection{The mean field approximation and Gaussian Ansatz for fluctuations}
We define the mean field by the quantum statistical average of the quantum fields and then introduce as in I, 
the Gaussian Ansatz for the density operator $\hat{\rho}$ with respect to fluctuations
\begin{eqnarray}
\phi_{c, i} (\v{r},t) &=& \langle  \hat{\phi}_i (\v{r},t) \rangle = \mbox{tr}(\hat{\rho} \hat{\phi}_i \rt), \\
\pi_{c, i}(\v{r},t) &=& \langle  \hat{\pi}_i (\v{r},t) \rangle.
\end{eqnarray}
$\tphii \rt $ and $ \tilde{\pi}_i \rt$ are given by
\begin{eqnarray}
\tilde{\phi}_i(\v{r},t) = \hat{\phi}_i(\v{r},t) - \phi_{c, i}(\v{r},t) \\
\tilde{\pi}_i(\v{r},t) = \hat{\pi}_i(\v{r},t) - \pi_{c, i}(\v{r},t)
\end{eqnarray}

\begin{eqnarray}
\langle \tilde{\phi}_i \rt \rangle  &=&0 \\
\langle \tilde{\phi}_i^3 \rt \rangle&=& 0
\end{eqnarray}
and
\begin{eqnarray}
\langle \left(  \sum_{i = 1}^N \tilde{\phi}_i^2(\v{r},t) \right)^2 \rangle
& = &3 \sum_{i=1}^N \langle \tilde{\phi}_i^2 (\v{r},t) \rangle^2 \nonumber \\
& & \qquad + \sum_{i \neq j} \left( 2\langle \tphi_i(\v{r},t) \tphi_j(\v{r},t) \rangle ^2 
+ \langle \tphi_i^2(\v{r},t)\rangle \langle \tphi_j^2(\v{r},t)\rangle \right) \nonumber \\
\end{eqnarray}
This Ansatz for the density operator implies a non-equilibrium generalization of the Hartree approximation in equilibrium.  
These shifted field operators obey the same equal-time commutation relations as the original field:
\begin{eqnarray}
\left[ \tilde{\phi}_i(\v{r},t),\tilde{\pi}_j(\v{r'},t) \right] &=& i \delta_{i,j} \delta (\v{r}-\v{r'}) \label{O(N):c.r.1}\\
\left[ \tilde{\phi}_i (\v{r},t),\tilde{\phi}_j (\v{r'},t) \right] &=& \left[ \tilde{\pi}_i (\v{r},t),\tilde{\pi}_j (\v{r'},t) \right]  = 0 \label{O(N):c.r.2}
\end{eqnarray}

\subsection{The Wigner functions}
To derive a kinetic equation for quantum fluctuation $\tphi_i$, we first define the particle creation and annihilation operators as
\begin{eqnarray}
\tilde{\phi}_i(\v{r},t) &=& \sum_{\vec{p}} \e^{i\v{p} \cdot \v{r}}
\frac{1}{ \sqrt{ 2\omega_{i,\v{p}} } } \left[ a_{i,\v{p}}(t) + a_{i,-\v{p}}^{\dagger} (t) \right] \\ 
\tilde{\pi}_i (\v{r},t) &=& i \sum_{\vec{p}} \e^{i\v{p} \cdot \v{r}}
\sqrt{\frac{\omega_{i,\v{p}}}{2}} \left[ a_{i,\v{p}}^{\dagger} (t) - a_{i,-\v{p}} (t) \right] 
\end{eqnarray}
with 
\begin{eqnarray}
\omega_{i,\p}=\sqrt{\p^2 + \mu_i^2 } \label{omega-mui}
\end{eqnarray}
The operators obey the usual equal-time commutation relations:
\begin{eqnarray}
\left[ a_{i,\v{p}}(t), a_{j,\v{p}'}^{\dagger} (t) \right] &=& \delta_{i,j} \delta_{\v{p},\v{p}'} 
\label{com1} \\ 
\left[ a_{i,\v{p}}(t), a_{j,\v{p}'} (t) \right] &=& 
 \left[ a_{i,\v{p}}^{\dagger}(t), a_{j,\v{p}'}^{\dagger} (t) \right] = 0
 \label{com2}
\end{eqnarray}
We introduce the Wigner functions of $O(N)$ model by
\be
     F_{ij}(\v{p},\v{k},t)  &=& \langle  a^{\dagger}_{i,\v{p}+\v{k}/2}(t) a_{j,\v{p}-\v{k}/2}(t) \rangle 
     \label{Fij} \\
     \bar{F}_{ij}(\v{p},\v{k},t) &=& \langle a_{i,-\v{p}-\v{k}/2}(t) a_{j,-\v{p}+\v{k}/2}^{\dagger}(t) \rangle 
     \label{Fbarij} \\
     \bar{G}_{ij}(\v{p},\v{k},t) &=&  \langle a^{\dagger}_{i,-\v{p}-\v{k}/2}(t) a^{\dagger}_{j,\v{p}-\v{k}/2}(t) \rangle \label{Gij}\\
     G_{ij}(\v{p},\v{k},t)  &=& \langle a_{i,-\v{p}-\v{k}/2}(t) a_{j,\v{p}-\v{k}/2}(t) \rangle \label{Gbarij}
\ee
These four Wigner functions are not independent but are related each other by the commutation
relations (\ref{com2}), e. g. 
\be\label{F-bar}
\bar{F}_{ij}(\v{p},\v{k},t)  = F_{ji}(-\v{p},\v{k},t) + \delta_{ij} \delta_{\k, 0} 
\ee
Other relations are given in Appendix A.
 
If we introduce a two-component notation of the operators
\be
A_{i,\p}(t) &=& \left( a_{i,\p}(t), \quad a^{\dagger}_{i,-\p} (t) \right) \quad \mbox{and} \quad A^{\dagger}_{i,\p}(t) = \left(
  \begin{array}{c}
	a^{\dagger}_{i,\p} (t)  \\
	a_{i,-\p} (t)   \\
  \end{array}
\right)
\ee
the Wigner functions may be written in a matrix form as 
\be
\v{W}_{ij} (\p,\k,t) &=& \la A^{\dg}_{i,\p +\k/2}(t) A_{j,\p - \k/2} \ra 
= \left(
  \begin{array}{cc}
    F_{ij}(\v{p},\v{k},t)   & \bar{G}_{ij}(\v{p},\v{k},t)    \\
    G_{ij}(\v{p},\v{k},t)  & \bar{F}_{ij}(\v{p},\v{k},t)   \\
  \end{array}
\right)
\ee
The Fourier transforms of the Wingner functions are defined by
\be
\v{w}_{ij} (\p,\r,t) = \sum_{\k} e^{-i \k \cdot \r} \v{W}_{ij} (\p,\k,t)
=
\left(
  \begin{array}{cc}
     f_{ij}(\v{p},\v{r},t)  & \bar{g}_{ij}(\v{p},\v{r},t)   \\
     g_{ij}(\v{p},\v{r},t)  & \bar{f}_{ij}(\v{p},\v{r},t)   \\
  \end{array}
\right)
\ee
where (\ref{F-bar}) implies 
\be\label{f-bar}
\bar{f}_{ij} (\v{p},\v{r},t) = f_{ji} (-\v{p},\v{r},t) + \delta_{ij} 
\ee

As we have noted in I, we consider $\mu_i$ in (\ref{omega-mui}) as free parameters which may be chosen as different from 
the mass parameter $m$ in the original Hamiltonian of $O(N)$ model. 
Generally, physical particle masses for interacting fields may differ from the mass parameters in the Hamiltonian due to the 
various kinds of dynamical effects.
Here we focus on time-evolution of the nonequilibrium system where the physical particle mass cannot be defined globally 
to take a fixed value.
Since a different choice of the mass parameter provides us with a different definition for ``particle excitations" of a nonequilibrium 
system, we should make proper use of the Wigner functions with an appropriately chosen mass parameter $\mu_i$ when 
we compare the results with observed particle distribution.

Possible instability of the system is signified by the appearance of negative value of mass square if we determine it self-consistently.
However, if we fix the mass parameters $\mu_i$, this instability, if exists, would manifest itself as an instability of the solution of 
the equations of motion for the condensate.

Suppose we choose a different particle ``mass" $\mu_i'$ to define the creation and annihilation operators:
\be
A'_{i,\p}(t) = (a'_{i,\p}(t), \quad a'^{\dg}_{i,-\p} (t))
\ee 
where
\begin{eqnarray}
\tilde{\phi}_i(\v{r},t) &=& \sum_{\v{p}} \e^{i\v{p} \cdot \v{r}}
\frac{1}{ \sqrt{ 2\omega'_{i,\v{p}} } } \left[ a'_{i,\v{p}}(t) + a'^{\dagger}_{i,-\v{p}} (t) \right] \\ 
\tilde{\pi}_i (\v{r},t) &=& i \sum_{\v{p}} \e^{i\v{p} \cdot \v{r}}
\sqrt{\frac{\omega'_{i,\v{p}}}{2}} \left[ a'^{\dagger}_{i,\v{p}} (t) - a'_{i,-\v{p}} (t) \right] 
\end{eqnarray}
with
\begin{eqnarray}
\omega'_{i,\v{p}} = \sqrt{\p^2 + \mu_i'^2}.
\end{eqnarray}
These new operators should also obey the commutation relations
\begin{eqnarray}
\left[ a'_{i,\v{p}}(t), a'^{\dagger} _{j,\v{p}'}(t) \right] &=& \delta_{i,j} \delta_{\v{p},\v{p}'} \\ 
\left[ a'_{i,\v{p}}(t), a'_{j,\v{p}'} (t) \right] &=& 
 \left[ a'^{\dagger} _{i,\v{p}}(t), a'^{\dagger} _{j,\v{p}'}(t) \right] = 0,
\end{eqnarray}
and they are related to the original ones by the Bogoliubov transformation:
\be
A'_{i,\p}(t) =A_{i,\p}(t) e^{\alpha_{i,\p} \tau_1 } 
\ee
where the real parameter $\alpha_{i,p}$ is given by
\begin{eqnarray}
\alpha_{i,p} = \frac{1}{2}\log \left( \frac{\omega'_{i,\p}}{\omega_{i,\p} } \right)
= \frac{1}{4}\log \left( \frac{\p^2+\mu_i'^2}{\p^2+\mu_i^2} \right)
\end{eqnarray}
and $\tau_1, \tau_2, \tau_3$ are the Pauli matrix
\begin{eqnarray}
\tau_1 = \left(
  \begin{array}{cc}
     0  & 1   \\
     1  & 0   \\
  \end{array}
\right) ,
\tau_2 = \left(
  \begin{array}{cc}
     0  & -i   \\
     i  & 0   \\
  \end{array}
\right) ,
\tau_3 = \left(
  \begin{array}{cc}
     1  & 0   \\
     0  & -1   \\
  \end{array}
\right) 
\end{eqnarray}
The relations between the new Wigner functions defined by the new creation and annihilation operators ($a'^{\dagger}_{i}, a_i$) and 
the original Wigner functions are described as the following matrix equation:
\begin{eqnarray}
\v{W}_{ij}'(\v{p},\v{k},t)
&=& \langle A'^{\dg}_{i,\p+ \frac{\k}{2} }(t) A'_{j,\p-\frac{\k}{2} } (t) \rangle \\
&=& e^{\tau_1\alpha_{i,\p+\frac{\k}{2}}} \langle A^{\dg}_{i,\p+ \frac{\k}{2}} (t) A_{j,\p-\frac{\k}{2}}(t) \rangle e^{ \tau_1\alpha_{j,\p-\frac{\k}{2}}} \\
&=& e^{ \tau_1 \alpha_{i,\p+\frac{\k}{2}}} \v{W}_{ij}(\v{p},\v{k},t) e^{ \tau_1 \alpha_{j,\p-\frac{\k}{2}}}
\end{eqnarray}

For small deviations of the mass parameters $\mu_i \rightarrow \mu_i +\delta \mu_i$ and $\mu_j \rightarrow \mu_j +\delta \mu_j$, 
the change of the Wigner functions will be written by
\begin{eqnarray}
\delta \v{W}_{ij}(\v{p},\v{k},t) &=& \tau_1 \v{W}_{ij}(\v{p},\v{k},t) \delta \alpha_{i,\v{p}+\v{k}/2} + \v{W}_{ij}(\v{p},\v{k},t) \tau_1 \delta \alpha_{j,\v{p}-\v{k}/2} \nonumber \\   
&=&  \frac{\tau_1 \v{W}_{ij}(\v{p},\v{k},t) \mu_i \delta \mu_i }{2((\v{p}+\v{k}/2)^2 + \mu_i^2)}
  + \frac{\v{W}_{ij}(\v{p},\v{k},t) \tau_1 \mu_j \delta \mu_j}{2((\v{p}-\v{k}/2)^2 + \mu_j^2)} 
\end{eqnarray}
\subsection{Equation of motion in the mean field approximation}
We obtain the equation of motion of the classical mean field $\phi_{c,i} \rt$ by taking the quantum statistical average of the Klein-Gordon equation of the scalar field of $\hphi_i$ (\ref{eq:ON-KG}).
With the Gaussian Ansatz for the density matrix, we obtain the equations of motion of the classical mean field $\phi_{c,i} \, (i=1,2, ... , N)$:
\begin{eqnarray}
\hspace{-0.5cm} (\Box+ m^2) \phi_{c, i} \rt = - \frac{\lambda}{3!} \left[ \phi_{c,i} \sum_{j=1}^N(\phi_{c,j}^2 + \langle \tphi_j^2 \rangle) + 2 \sum_{j=1}^N \phi_{c,j} \langle\tphi_i \tphi_j \rangle 
 \right]-h_i \label{NLKGn}
\end{eqnarray}
As we have done so in I, we may call these classical equations {\it non-linear Klein-Gordon equations} 
of $O(N)$ sigma model corresponding to the non-linear Sch\"odinger equation (or Gross-Pitaevskii equation) in the theory of 
the Bose-Einstein condensates. 
The non-linearity originates from the self-interaction of the classical fields $\phi_{c,i}$, the interaction between $\phi_{c,i}$ and $\phi_{c,j}$, 
and also from the interaction with the fluctuations $\langle \tphi_i \rt \tphi_j \rt \rangle$ which implicitly depends on $\phi_{c,i}$.  
The fluctuations $\langle \tphi_i \rt \tphi_j \rt \rangle$ arise from ``particle excitations" because of the following expression in terms of 
the Wigner functions as
\begin{eqnarray}
\langle \tphi_i \rt \tphi_j \rt \rangle 
&=&  \sum_{\v{p},\v{p}'} \frac{e^{i(-\v{p}+\v{p}') \cdot \v{r}}}{\sqrt{2\omega_{i,\v{p}}}\sqrt{2\omega_{j,\v{p}'}}}
 \langle (a_{i,-\v{p}}+a^{\dagger}_{i,\v{p}})(a_{j,\v{p}'}+a^{\dagger}_{j,-\v{p}'}) \rangle \nonumber \\  
&=& \sum_{\v{p},\v{p}'} \frac{e^{i(-\v{p}+\v{p}') \cdot \v{r} } \tr \left[ (1+\tau_1)\v{W_{ij}}(\frac{\v{p}+\v{p}'}{2} , \v{p}-\v{p}' , t) \right]
 }{\sqrt{2\omega_{i,\v{p}}}\sqrt{2\omega_{j,\v{p}'}}}           
\label{tphiij}
\end{eqnarray}
Thus, the time-evolution of the classical mean field $\phi_{c,i}$ is coupled with that of the Wigner functions.

The equation of motion of the Wigner functions may be obtained from the commutators of bilinear forms of $a_{i,\v{p}}$ and 
$a^{\dagger}_{j,\v{p}}$  with the mean field Hamiltonian defined by
\begin{eqnarray}
H_{\text{mf}} &=& \int d\v{r} \frac{1}{2}
   \left[ 
   \sum_i 
   	\left( \tilde{\pi_i}^2 + 
     (\nabla \tilde{\phi_i})^2 + m^2 \tilde{\phi_i}^2 \right)
       + \sum_{ij} \Pi_{ij}(\v{r},t) \tilde{\phi_i} \tilde{\phi_j} \right] 
       \nonumber \\ 
       &=& \int d\v{r} \frac{1}{2}
   \left[ 
   \sum_i 
   	\left( \tilde{\pi_i}^2 + 
     (\nabla \tilde{\phi_i})^2 + \mu_i^2 \tilde{\phi_i}^2 \right)
       + \sum_{ij} \Delta \Pi_{ij}(\v{r},t) \tilde{\phi_i} \tilde{\phi_j} \right]
\end{eqnarray}
where
\begin{eqnarray}
\Pi_{ij} &=& \frac{\lambda}{6} \delta_{ij} \sum_{k=1}^N \left( \phi_{c,k}^2 + \langle \tphi_k^2 \rangle \right) +
 \frac{\lambda}{3} \left( \phi_{c,i} \phi_{c,j} + \langle \tphi_i \tphi_j \rangle \right)   \label{Pi_ij} 
\end{eqnarray}
and 
\begin{eqnarray}
\Delta \Pi_{ij}&=&\Pi_{ij} \rt + (m^2 - \mu_i^2) \delta_{ij} \label{DeltaPi_ij}
\end{eqnarray}
which implies that the mass parameters are given by
\be
\mu_i^2 = m^2  + \frac{\lambda}{3} \left( \phi_{c,i}^2 + \langle \tphi_i^2 \rangle \right) 
 + \frac{\lambda}{6} \sum_{j=1}^N \left( \phi_{c,j}^2 + \langle \tphi_j^2 \rangle \right)
 \label{mu_i}
\ee

The momentum representation of this mean field Hamiltonian can be written as
\begin{eqnarray}
\hspace{-1cm}
H_{\text{mf}}=\sum_{\v{p},i} \omega_{i,\v{p}} a^{\dagger}_{i,\v{p}} a_{i,\v{p}} + \frac{1}{2} \sum_{\v{p},\v{q},i,j} \Delta \Pi_{ij,\v{q}} \cdot 
    \frac{(a_{i,\v{p}}+a^{\dagger}_{i,-\v{p}})
    (a_{j,-\v{p}-\v{q}}+a^{\dagger}_{j,\v{p}+\v{q}})}
    {\sqrt{2\omega_{i,\v{p}}}\sqrt{2\omega_{j,\v{p}+\v{q}}}}
\end{eqnarray}
where
\begin{eqnarray}
\Delta\Pi_{ij,\v{q}}(t) &=& \int d\v{r} e^{-i\v{p} \cdot \v{r}}
                (\Pi_{ij}(\v{r},t) + ( m^2 - \mu_i^2) \delta_{ij}) \nonumber \\
                &=& \Pi_{ij,\v{q}}(t) + (m^2 - \mu_i^2) \delta_{ij} \delta_{\v{q},0}
\end{eqnarray}
In the Heisenberg picture, time-derivative of operators are calculated from the commutator of the Hamiltonian. 
With the Gaussian Ansatz for the density operator, the average of the commutator with the original Hamiltonian is equivalent to that 
with the mean field Hamiltonian $H_{\text{mf}}$:
\begin{eqnarray}
i \dot{\v{W}}_{ij}(\v{p},\v{k},t) = \left<
\left[ A^{\dg}_{i,\v{p}+\frac{\v{k}}{2}} A_{j,\v{p}-\frac{\v{k}}{2}} 
 , H \right] \right>
= \left<
\left[ A^{\dg}_{i,\v{p}+\frac{\v{k}}{2}} A_{j,\v{p}-\frac{\v{k}}{2}} 
 , H_{\text{mf}} \right] \right>
\end{eqnarray}

Therefore, we obtain the equation of motion of the Wigner function $\v{W}_{ij} (\v{p},\v{k},t)$.
These equations can be written in a matrix equation as:
\be
\hspace{-1.3cm}
i \frac{\partial}{\partial t} \v{W}_{ij} ( \bp, \bk, t ) & = &  
- \omega_{i,\bp + \bk/2} \tau_3 \v{W}_{ij}( \bp, \bk, t )  + \omega_{j,\bp - \bk/2}  \v{W}_{ij} ( \bp, \bk, t ) \tau_3
\nonumber \\
& & \quad  - \sum_{l,\bq} \DPi_{il\bq}
\frac{ \tau_3(1+ \tau_1) \v{W}_{lj} ( \bp + \bq/2, \bk + \bq, t )  }
{\sqrt{2 \omega_{i,\bp + \bk/2}}\sqrt{2 \omega_{l,\bp+ \bk/2 + \bq }}}
\nonumber \\
& &
\qquad  + \sum_{l,\bq} 
\frac{\v{W}_{il}( \bp - \bq/2, \bk + \bq, t )(1+ \tau_1)\tau_3 }
{\sqrt{2 \omega_{l,\bp - \bk/2}}\sqrt{2\omega_{j,\bp - \bk/2 - \bq }}}
\DPi_{lj,\bq}
 \label{eomWij}
\ee 

Eq. (\ref{eomWij}) and the nonlinear Klein-Gordon equations (\ref{NLKGn}) form a closed system of coupled differential equations.
We note that the effect of the exthernal field $h_i$, only appears in modification of the nonliear Klein-Gordon equations.

\section{Kinetic equations for slowly varying system: the Vlasov equations of $O(N)$ model}
In this section we derive kinetic equation of $O(N)$ model in the long wavelength limit. 
Here we assume inhomogeneity of the system is due to the long wavelength fluctuation $k,q, \ll p$. 
We use the following approximations:
\begin{eqnarray}
\omega_{i,\v{p} \pm \v{k}/2} &\simeq& \omega_{i,\v{p}} \pm \frac{\v{p} \cdot \v{k}}{2 \omega_{i,\v{p}}} \\
\omega_{i,\v{p} + \v{k}/2} \pm \omega_{j,\v{p} - \v{k}/2} &\simeq& \left( \omega_{i,\v{p}} \pm \omega_{j,\v{p}} \right) 
+ \frac{ \v{p} \cdot \v{k} }{2} \left( \frac{1}{\omega_{i,\v{p}}} \mp \frac{1}{\omega_{j,\v{p}}} \right)
\end{eqnarray}
and
\begin{eqnarray}
\frac{1}{\sqrt{\omega_{i,\v{p} \pm \v{k}/2 }}\sqrt{\omega_{j,\v{p} \pm (\v{k}/2 +\v{q}) }}}  \simeq \frac{1}{\sqrt{\omega_{i,\v{p}} \omega_{j,\v{p}}}} 
\left( 1 \mp \frac{\v{p} \cdot \v{k}}{4 \omega^2_{i,\v{p}}} \mp \frac{ \v{p} \cdot ( \v{k} + 2 \v{q} )}{ 4  \omega^2_{j,\v{p}}} \right)
\end{eqnarray}
We also use the Taylor series expansion of the Wigner functions:
\begin{eqnarray}
\v{W}_{ij}(\v{p} \pm, \v{q}/2, \v{k} + \v{q},t) &\simeq& \v{W}(\v{p}, \v{k}+\v{q},t) \pm \frac{1}{2} \v{q} \cdot \nabla _{\v{p}} \v{W}_{ij}(\v{p}, \v{k}+\v{q},t) 
\end{eqnarray}
Then the equation of motion of the Wigner function $\v{w}_{ij}(\p,\r,t)$ becomes
\begin{eqnarray}
 \dot{\v{w}}_{ij} (\p,\r,t)&=& i(\w_{\p,i} \tau_3 \v{w}_{ij}(\p,\r,t)-\w_{\p,j} \v{w}_{ij}(\p,\r,t) \tau_3 )  \nonumber \\
&&\hspace{-1.5cm} -\left( \frac{1}{\w_{i,\p}^2}+\frac{1}{\w_{j,\p}^2} \right) \frac{\p}{2} \cdot \nabla (\tau_3 \v{w}_{ij}(\p,\r,t) -\v{w}_{ij}(\p,\r,t) \tau_3) \nonumber \\
&&\hspace{-1.5cm} + \tau_3 (1+\tau_1) 
\sum_l \Big[ 
i U_{il}(\r,t) \v{w}_{lj}(\p,\r,t)
 + \frac14 \left( \frac1{\w_{i,\p}^2} + \frac1{\w_{l,\p}^2}  \right) \p \cdot \nabla (U_{il}(\r,t) \v{w}_{lj}(\p,\r,t)) \nonumber \\
&&\hspace{-1.5cm} -\frac{\p}{2\w_{l,\p}^2} \cdot(\nabla U_{il}(\r,t))\v{w}_{lj}(\p,\r,t) + (\nabla U_{il}(\r,t))\cdot \nabla_{\p} \v{w}_{lj}(\p,\r,t) 
 \Big] \nonumber \\
&&\hspace{-1.5cm} + \sum_l \Big[ 
-i U_{lj}(\r,t) \v{w}_{il}(\p,\r,t)
 + \frac14 \left( \frac1{\w_{l,\p}^2} + \frac1{\w_{j,\p}^2}  \right) \p \cdot \nabla (U_{lj}(\r,t) \v{w}_{il}(\p,\r,t)) \nonumber \\
&&\hspace{-1.5cm} -\frac{\p}{2\w_{j,\p}^2} \cdot(\nabla U_{lj}(\r,t))\v{w}_{il}(\p,\r,t) + (\nabla U_{lj}(\r,t))\cdot \nabla_{\p} \v{w}_{il}(\p,\r,t) 
 \Big] (1+\tau_1) \tau_3 \label{eomwij}
\end{eqnarray}
where we have defined the generalized mean field potential $U_{ij}$ as 
\begin{eqnarray}
 U_{ij}(\v{r},t) = \frac{\Delta \Pi_{ij}(\v{r},t)}{2\sqrt{\omega_{i,\v{p}} \omega_{j,\v{p}}}} \label{Uij}
\end{eqnarray}
In the long wavelength approximation, we also have
\begin{eqnarray}
\langle \tilde{\phi}_i(\v{r},t) \tilde{\phi}_j(\v{r},t)  \rangle  \simeq 
 \sum_{\v{p}} \frac{\mbox{tr}\left[ (1+\tau_1) \v{w}_{ij}(\p,\r,t) \right]}{2 \sqrt{ \omega_{i,\v{p}}  \omega_{j,\v{p}}}}
\end{eqnarray}
which appears in both the self energy term $\Pi_{ij}$,(\ref{Pi_ij}) and the generalized mean field 
potential $U_{ij}$ (\ref{Uij}) and in the non-linear Klein-Gordon equation for the condensate 
$\phi_{c,i}$ (\ref{NLKGn}).

These equations constitute the generalization of the Vlasov equation we derived for one component scalar field in I.   
The equations of each components of $\v{w}_{ij}$ are shown explicitly in Appendix. 
They have rather complex structure. 
The diagonal components ($i = j$), of these equation can be rewritten in the form similar to the classical Vlasov equation.  
In fact, we extract the terms which contain the diagonal components of the Wigner function.   
Writing  $f_{ii} = f_i$, $U_{ii} = U_i$, we obtain  
\begin{eqnarray}\label{vlasov}
\dot{f}_i (\v{p},\v{r},t)  & = & -  \frac{1}{\omega_{i,\v{p}}} \left( 
1- \frac{U_i (\v{r},t)}{2\omega_{i,\v{p}}}  \right) \v{p} \cdot \nabla _{\v{r}}  f_i (\v{p},\v{r},t) 
+ \nabla_{\v{r}} U_i (\v{r},t) \cdot \nabla_{\v{p}} f_i (\v{p},\v{r},t)
\nonumber \\
\nonumber \\
& & \qquad \qquad + I_f + I_g  + I_{\bar{g}} 
\end{eqnarray}
where $I'_f$ contains the terms proportional to the off-diaginal components of 
$f_{ij}$ 
\be
I_f = &&  
\sum_{j \ne i} 
\Biggl[ 
i \biggl(U_{ij}(\v{r},t) f_{ji}(\v{p},\v{r},t) - f_{ij}(\v{p},\v{r},t) U_{ji}(\v{r},t)  \biggr)
\nonumber \\
&& \quad
+ \frac{1}{4} \left( \frac{1}{\omega^2_{i,\v{p}}} - \frac{1}{\omega^2_{j,\v{p}}} \right) 
 \biggl(  \v{p} \cdot  \nabla _{\v{r}} U_{ij}(\v{r},t)  f_{ji}(\v{p},\v{r},t) 
+ \v{p} \cdot  \nabla _{\v{r}} f_{ij}(\v{p},\v{r},t) U_{ji}(\v{r},t)  \biggr)
 \nonumber \\
&& \quad
+ \frac{1}{4} \left( \frac{1}{\omega^2_{i,\v{p}}} + \frac{1}{\omega^2_{j,\v{p}}} \right) 
\biggl( U_{ij} ( \v{r}, t ) \v{p} \cdot  \nabla _{\v{r}}  f_{ji} (\v{p},\v{r},t)  
+ \v{p} \cdot  \nabla _{\v{r}} f_{ij}(\v{p},\v{r},t) U_{ji} (\v{r},t)  \biggr)
\nonumber \\
&& \quad
+ \frac{1}{2} \biggl( \nabla_{\v{r}} U_{ij}(\v{r},t) \cdot \nabla_{\v{p}} f_{ji}(\v{p},\v{r},t)  
+  \nabla_{\v{p}} f_{ij}(\v{p},\v{r},t) \cdot  \nabla_{\v{r}} U_{ji}(\v{r},t)  \biggr)
\Biggr]
\ee
while $I_g$ and $I_{\bar{g}}$ are the terms linear in $g_{ij}$ and $\bar{g}_{ij}$, respectively:
\be
I_g = && \sum_j  \Biggl[ 
i U_{ij}(\v{r},t) g_{ji}(\v{p},\v{r},t) 
- \frac{1}{4} \left( \frac{1}{\omega^2_{i,\v{p}}} - \frac{1}{\omega^2_{j,\v{p}}} \right) \v{p} \cdot 
 \nabla _{\v{r}} U_{ij}(\v{r},t) g_{ji}(\v{p},\v{r},t) \nonumber \\
 && \hspace{2cm}  
- \frac{1}{4} \left( \frac{1}{\omega^2_{i,\v{p}}} + \frac{1}{\omega^2_{j,\v{p}}} \right) U_{ij}(\v{r},t) \v{p} 
\cdot  \nabla _{\v{r}}  g_{ji}(\v{p},\v{r},t)  \nonumber \\
&& \hspace{4cm}
- \frac{1}{2} \nabla_{\v{r}} U_{ij}(\v{r},t) \cdot \nabla_{\v{p}} g_{ji}(\v{p},\v{r},t) \Biggr]
 \\
I_{\bar{g}} = && \sum_j \Biggl[ 
- i \bar{g}_{ij}(\v{p},\v{r},t)  U_{ji}(\v{r},t) 
+ \frac{1}{4} \left( \frac{1}{\omega^2_{i,\v{p}}} - \frac{1}{\omega^2_{j,\v{p}}} \right) 
\bar{g}_{ij}(\v{p},\v{r},t) \v{p} \cdot \nabla _{\v{r}} U_{ji}(\v{r},t)  \nonumber \\
&& \hspace{2cm}  
+ \frac{1}{4} \left( \frac{1}{\omega^2_{i,\v{p}}} + \frac{1}{\omega^2_{j,\v{p}}} \right) 
\v{p} \cdot \nabla _{\v{r}}  \bar{g}_{ij}(\v{p},\v{r},t) \cdot U_{ji}(\v{r},t)  
  \nonumber \\
&& \hspace{4cm}
+ \frac{1}{2}  \nabla_{\v{p}} \bar{g}_{ij}(\v{p},\v{r},t)  \cdot \nabla_{\v{r}} U_{ji}(\v{r},t)
\Biggr]
\ee
Introducing the quasi-particle energy of the i-th species by 
\begin{equation}
\varepsilon_i (\v{p}, \v{r}, t )  = \omega_{i,\v{p}} + U_i (\v{r},t) 
\end{equation}
the first two terms on the right hand side of (\ref{vlasov}) takes a form of the Landau kinetic 
equation:
\begin{eqnarray} 
\dot{f}_i (\v{p},\v{r},t)  & = & 
-\nabla_{\v{p}} \varepsilon_i (\v{p}, \v{r}, t )  \cdot \nabla _{\v{r}}  f_i (\v{p},\v{r}, t) 
+ \nabla_{\v{r}} \varepsilon_i (\v{p}, \v{r} , t) \cdot \nabla_{\v{p}} f_i (\v{p},\v{r}, t) + \cdots 
\nonumber \\
\end{eqnarray}
where the first term of the right is the quasi-particle drift term with drift velocity given by
\begin{equation}
\nabla_{\v{p}} \varepsilon_i (\v{p}, \v{r}, t ) = \frac{1}{\omega_{i,\v{p}}}
\left( 1- \frac{U_i (\v{r},t ) }{2 \omega_{i,\v{p}}} \right) \v{p} \cdot \nabla _{\v{r}} 
\end{equation}

The other equations for the time derivative of the off-diagonal components $f_{ij}$ as well as 
$g_{ij}$ and ${\bar g}_{ij}$ have no classical counter-parts.   

%section 4
\section{Statistical equilibrium and the gap equations}
In this section, we discuss equilibrium solution of coupled kinetic equation for $O(N)$ model in spatially uniform system.
 
Taking $h_i=\epsilon \delta_{1i}$, we assume that only one component of the meson field 
$\phi_1$ has non-vanishing expectation value in equilibrium:
\begin{eqnarray}\label{phi0i}
\phi_{c, i} = 
\begin{cases}
\phi_{\rm eq}	&	 (i=1),   \\
0		&	 (i \ge 2) 
\end{cases}
\end{eqnarray}
Inserting this condition into the non-linear Klein-Gordon equations (\ref{NLKGn}), we have
\be
m^2 \phi_{\rm eq} & = & 	- \frac{\lambda }{3!} 
\left( 
\phi_{\rm eq}^2 + 2 \langle \tphi_1^2 \rangle_{\rm eq.} + \sum_{j=1}^N \langle \tphi_j^2 \rangle_{\rm eq.} 
\right) \phi_{\rm eq}
- \epsilon ~~.
\label{staticNLKG} 
\ee
We also assume 
\be
\langle \tphi_1 \tphi_j \rangle_{\rm eq}  =  0 
\ee
for $ j=2, \cdots , N$. 
In the chiral limit $\epsilon = 0$, (\ref{staticNLKG}) has non-vanishing solutions for 
$\phi_{\rm eq} (\ne 0) $ only at low temperatures  
which satisfy
\be
m^2  &=& - \frac{\lambda }{3!} 
\left( 
\phi_{\rm eq.}^2 + 2 \langle \tphi_1^2 \rangle_{\rm eq.} + \sum_{j=1}^N \langle \tphi_j^2 \rangle_{\rm eq.} 
\right)  
\label{StaticEq_lowT}
\ee

In order to ensure that the time derivative of the Wigner functions vanishes,  
we furthermore impose the condition that the off-diagonal components of the self-energy 
$\Delta \Pi_{ij}$ or the mean field potential $U_{ij}$ vanish in uniform equilibrium 
\be
\Delta \Pi^{\rm eq.}_{ij} &=& \Delta \Pi^{\rm eq.}_{ji} = 
U^{\rm eq.}_{ij}  = U^{\rm eq.}_{ji} = 0 \label{U0ij}
\ee
and so do the all off-diagional components of the Winger functions: 
\be
f^{\rm eq.}_{ij} &=& g^{\rm eq.}_{ij} = \bar{g}^{\rm eq.}_{ij}  =0 \qquad \qquad (i \neq j) \,
\quad 
\ee
Finally we set the diagonal components of the mean field potential also vanish: 
\be
\Delta \Pi^{\rm eq.}_{ii} = U^{\rm eq.}_{ii} = 0 \label{U0ii}
\ee
so that the mass parameter $\mu_i$ are chosen as
\be
\mu_i^2 = m^2 + \Pi^{\rm eq.}_{ii}  \label{mu_eq}
\ee
This last procedure eliminates the diagonal components of the $g_{ij}$ and $\bar{g}_{ij}$.

Now the only non-vanishing components of the Wigner functions are diagonal components
of $f_{ij}$
\be
f_{ij} &=& \bar{f}_{ij} = \foi \delta_{ij} 
\ee
 for which we take Bose distribution function: 
\be
\foi = \frac{1}{\e^{ \beta \omega_{i,\p} }-1} 
\label{foi}
\ee
where $\beta=1/k_B T$ is the inverse temperature. 

The conditions (\ref{phi0i}), (\ref{U0ij}) and (\ref{mu_eq}) imply that the mass parameters 
should be chosen as

\be 
\mu_i^2 &=&m^2+  \frac{\lambda}{6} \left[  ( 1+ 2 \delta_{i,1}) \phi_{\rm eq}^2
+ 2 \langle \tilde{\phi}_i^2 \rangle _{\rm eq.} +\sum_{j=1}^N \langle \tilde{\phi}_j^2 \rangle _{\rm eq.} \right]  \label{ONgapeq2} 
\ee
where the thermal fluctuations of quantum fields are given by the relation (\ref{tphiij}) as
\be
\langle \tilde{\phi}_i^2 \rangle _{\rm eq.} = \sum_{\p} \frac{1}{\omega_{i,\p}} \foi
= \frac{1}{(2 \pi)^3} \int  \frac{d \bp}{\omega_\bp} \frac{1}{e^{\omega_{i, \p} \beta} - 1}  
 =  \frac{\beta^{-2}}{2\pi^2}  I^{(2)}_- ( \mu_i \beta)  
 \label{ThFluc}
\ee
where the dimensionless function $I^{(2)}_- ( x ) $ is given by  \cite{MM08}
\begin{eqnarray}
I^{(2)}_- ( x )  & \equiv & \int_0^\infty \frac{k^2 d k}{\sqrt{k^2 + x^2}} \frac{1}{e^{\sqrt{k^2 + x^2}} - 1}
\label{integral} \\
& = & \frac{\pi^2}{6} - \frac{\pi}{2}x - \frac{1}{4}x^2 \ln \frac{x}{4\pi} +
\left( \frac{1}{8} - \frac{1}{4} \gamma \right) x^2 - \frac{\zeta (3)}{32 \pi} x^4 + {\cal O}  ( x^6)
\label{I2} 
\end{eqnarray}
where $\gamma = 0.57721 \cdots$ is Euler's number.  
We call these equations the gap equations of the $O(N)$ model. 

Although, by setting $h_i=\epsilon \delta_{1i}$ and choosing (\ref{phi0i}), we have committed breaking 
the original $O(N)$ symmetry in the direction of $\phi_1$ field, we still have an $O(N-1)$ symmetry with
respect to the rotions of $(\phi_2, \phi_3, \cdots, \phi_N)$.   
We still have degeneracy in the mass parameters $\mu_2=\mu_3=\cdots=\mu_N$. 
This implies that the original gap equations consisting of $N$ equations are reduced into 
the following two coupled equations:
\be
\mu_1^2 & = & m^2 +  
\frac{\lambda}{2} \phi_{\rm eq}^2 + \frac{\lambda}{2} \langle \tilde{\phi}_1^2 \rangle _{\rm eq.} 
+  \frac{\lambda}{6} (N-1)  \langle \tilde{\phi}_2^2 \rangle _{\rm eq.} 
\label{gap1}
\\
\mu_2^2 & = & m^2 + 
 \frac{\lambda}{6}  \phi_{\rm eq}^2 + \frac{\lambda}{6}  \langle \tilde{\phi}_1^2 \rangle _{\rm eq.} 
+  \frac{\lambda}{6} (N + 1) \langle \tilde{\phi}_2^2 \rangle _{\rm eq.}
\label{gap2}
\ee
and the static Klein-Gordon equation (\ref{staticNLKG}) reduces to
\be
m^2 \phi_{\rm eq} & = & 	- \frac{\lambda }{6} 
\left( 
\phi_{\rm eq}^2 + 3 \langle \tphi_1^2 \rangle_{\rm eq.} + (N-1) \langle \tphi_2^2 \rangle_{\rm eq.} 
\right) \phi_{\rm eq}
- \epsilon ~~.
\label{rStaticKG} 
\ee

The reduced gap equations (\ref{gap1}) - (\ref{gap2}) together with the reduced static Klein-Gordon 
equation (\ref{rStaticKG}) determine the values of $\mu_1$, $\mu_2$ and $\phi_{\rm eq.}$ 
at given temperature self-consistently.  
As we also noted in I,  our gap equations are identical to the variational conditions to determine 
the mass parameters in the CJT composite operator effective potential in the one-loop 
approximation\cite{A-C97} if we ignore the renormalization effect due to the divergent vacuum
polarization which we have dropped out in evaluating the Wigner functions.  

%%subsection 4.1
\subsection{Exact chiral limit $(\ep = 0)$}
We first show numerical results of the temperature dependence of the solutions of the gap 
equations (\ref{gap1}) - (\ref{gap2}) in the absence of symmetry breaking external field 
($ \epsilon = 0 $).   
The results are plotted in Fig. 1.

\begin{figure}[h]
\begin{center}
\includegraphics[scale=1.]{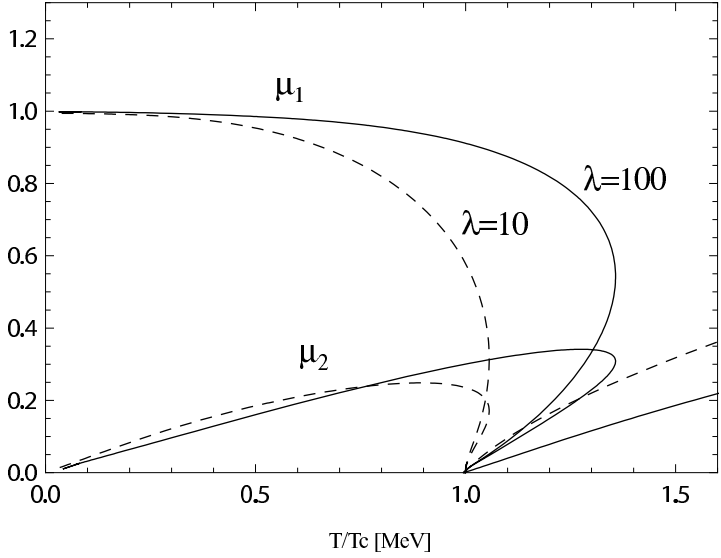}%GapEq_CL.eps}
\end{center}
\caption{Temperature dependence of the solutions of the gap equations for the scaled sigma mass $\mu_1/\mu_0$ and 
the scaled pion mass $\mu_2/\mu_0$ as a function of of $t=T/T_c$.  }
\label{fig1}
\end{figure}

In the low temperature phase where $\phi_{\rm eq.} \ne 0$, the gap equations are solved with 
the equation (\ref{StaticEq_lowT}) obtained from the static Klein-Gordon equation or
its reduced form
\be
m^2 = - \frac{\lambda }{6} 
\left( 
\phi_{\rm eq}^2 + 3 \langle \tphi_1^2 \rangle_{\rm eq.} + (N-1) \langle \tphi_2^2 \rangle_{\rm eq.} 
\right) 
\ee

Since all the thermal fluctuations $\langle \tilde{\phi}_i^2 \rangle _{\rm eq.}$ in the gap equations 
vanish at zero temperature, all mass parameters vanish except for $\mu_1$ which is determined by
\be
\mu_1^2 (T = 0) \equiv  \mu_0^2 = m^2 + \frac{\lambda}{2} \phi_0^2 = - 2 m^2
\ee 
where we have used the relation 
\be
m^2 = - \frac{\lambda}{3!} \phi_0^2
\ee
to determine the vacuum condensate $\phi_0$. 
In Fig. 1, the solution of the gap equations are scaled by the sigma meson mass $\mu_0$ at $T=0$ 
for $\hat{\mu}_1=\mu_1/\mu_0$ and $\hat{\mu}_2=\mu_2/\mu_0$ as a function of $t=T/T_c$ in the case 
of $N=4$. 
 
The two mass parameters $\mu_1, \mu_2$ and the condensate amplitude 
$\phi_{\rm eq.}$ all vanish at ``critical" temperature $T_c$ given by 
\be
(k_B T_c)^2 = - \frac{72}{(N+2)\lambda} m^2 
\ee
which is obtained by setting $\mu_1 = \mu_2 = \phi_{\bf eq.} = 0$ in equations 
(\ref{gap1}) and (\ref{gap2}) and using the leading term in the expansion (\ref{I2})
which gives $\langle \tilde{\phi}_1^2 \rangle _{\rm eq.} = 
\langle \tilde{\phi}_1^2 \rangle _{\rm eq.} = \beta^2 / 12$. 

In the high temperature phase,  the condensate vanishes $\phi_{\rm eq.} =0$ and 
all the mass paraeter become degenerate with
\be
\mu_1^2 = \mu_2^2 = m^2 + \frac{\lambda}{6} (N+2) \fla _{\rm eq.} 
\ee
which may be obtained from either (\ref{gap1}) or (\ref{gap2}) by setting 
$\phi_{\rm eq.} = 0$, $\mu_1 = \mu_2$, and
$\langle \tilde{\phi}_1^2 \rangle _{\rm eq.} = \langle \tilde{\phi}_2^2 \rangle _{\rm eq.}$.

The phase transition in this model becomes of first order showing a typical hysteresis behavior.
This behavior has been known as a generic symptom of the mean field approximation with the $O(N)$ sigma models 
with $\phi^4$ interactions \cite{BG77, RM98, CH98, Pet99, LR00}. 
It has been shown, however,  that the transition becomes second order if one includes the effect of the fluctuations 
in the two-loop approximation \cite{Chi00} based on the optimal perturbation method developed in \cite{CH98}.
The second order phase transition has been obtained also by the Nambu-Jona-Lasinio model\cite{CleymansKociifmmodeScadron1989,Bili'cCleymansScadron1995}.
Although, in the chiral limit, the pion mass $\mu_2$ is expected to vanish due to the Goldstone theorem, 
$\mu_2$ has non-vanishing values in the low temperature phase. 
This apparent violation of the Goldstone theorem always occurs in the mean field approximation.
With a resummation method known as an optimized perturbation theory at finite temperature developed in $O(N)$ $\phi^4$ 
theory, the Goldstone theorem is satisfied for arbitrary $N$\cite{CH98}. 
In our framework, the missing Goldstone theorem can be retrieved as the collective excitation mode 
\cite{Okopinska1996,DSM96,NGP97,TVM00} as we shall show 
in the next section.  

%% subsection 4.1
\subsection{With symmetry breaking $(\ep \neq 0)$}

In the presence of the explicit symmetry breaking ($\epsilon \ne 0$) the mass parameters 
$\mu_i$ for $i \ne 1$ also takes non-vanishing value even at zero temperature.  
This is seen by setting $\langle \tilde{\phi}_i^2 \rangle _{\rm eq.} = 0$ for all 
components of meson fields in (\ref{StaticEq_lowT}) which yields
\be
m^2 +  \frac{\lambda}{6} \phi_0^2 = \frac{\epsilon}{\phi_0}
\ee
where $\phi_0$ is the amplitude of the vacuum condensate.
Inserting this relation in (\ref{gap2}) gives 
\be
\mu_2^2 = \frac{\epsilon}{\phi_0}
\ee
at zero temperature. 
In the case of $N=4$ we may interpret this mass as the degenerate mass 
$m_\pi$ of three pions $(\pi_+, \pi_0, \pi_-)$ which form an isovector field. 
We thus set
\be\label{m_pi}
\epsilon = m_\pi^2 \phi_0
\ee
as a condition to choose the symmetry breaking parameter $\epsilon$.
One can also show that the equation of motion (\ref{eq:ON-KG}) 
imply that there is a partially conserved "axial vector" current 
\be
\partial_\mu ( \hphi_1 \partial^\mu \hphi_i  - \hphi_i \partial^\mu \hphi_1  ) = \epsilon \hphi_1  
\ee
Sandwitching this operator relation by the vacuum state and the single charged pion state
we obtain the relation which connect the vacuum expectation value of $\sigma$ field $\phi_1$,
namely $\phi_0$, to the charged pion decay constant $f_\pi$,
\be\label{f_pi}
\phi_0 = f_\pi %= 93 ~{\rm MeV}
\ee
The $\sigma$-meson mass in the vacuum is determined by (\ref{gap1}) at zero temperature 
\be\label{m_sigma}
m_\sigma^2 = m^2 + \frac{\lambda}{2} \phi_0^2 = m^2 + \frac{\lambda}{2} f_\pi^2
\ee

We use three conditions (\ref{m_pi}), (\ref{f_pi}) and (\ref{m_sigma}) to determine three
parameters $m^2$, $\lambda$, and $\epsilon$ in our model Hamiltonian and solve the 
gap equations (\ref{gap1}), (\ref{gap2}) and (\ref{rStaticKG}) to determine
the temperature dependence of $\mu_1$, $\mu_2$ and $\phi_{\rm eq.}$.    
Numerical values of the parameters are determined by these conditions to give 
$m_\sigma = \mu_1 = 500$MeV, $m_\pi = \mu_2 = 140$MeV,  $f_\pi = \phi_0 = 94$MeV 
at zero temperature by tuning $\lambda = 78.2$. 
We show numerical result computed with these values of parameters is shown in Fig. 2.  
It is seen that the transition is smoothed out with no trace of the first order transition.   

\begin{figure}[h]
\begin{center}
\includegraphics[scale=1.]{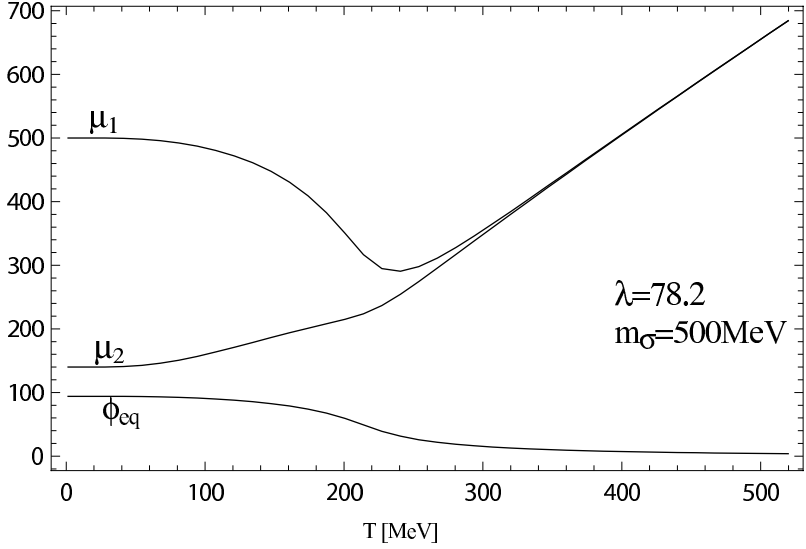}%GapEq_SSB.eps}
\end{center}
\caption{Temperature dependence of the solutions of the gap equations with explicit symmetry 
breaking. }
\label{fig2}
\end{figure}

%\newpage

\section{Dispersion relations of the collective excitations}
In this section we compute dispersion relation of the collective excitation in the system near equilibrium in $O(N)$ model.  We have shown in I that the collective mode appears in the low temperature phase
as a meson mode with small effective mass as a coupling of the mesonic excitation with the
quasi-particle excitations.  In the $O(N)$ sigma model there are $N$ different kind of quasi-particle 
excitations in the system in equilibrium.   We will show that the fluctuation of the condensate 
amplitude $\phi_{c, 1}$ couple with the fluctuation of the diagonal components of the
Wigner function $f_{ii}$, $g_{ii}$ and $\bar{g}_{ii}$ and generate the collective mode of the sigma
meson type, while the fluctuations of the condensate perpendicular to the condensate $\phi_{c, i}$ 
with $i \ge 2$ couple with the off-diagonal components of the Wigner functions and 
generate pionic collective modes, which constitute the missing Nambu-Goldstone modes  
in the chiral limit.   

We first assume that each of the meson field and the distribution functions consists of 
uniform equilibrium term and a small fluctuations around it:
\begin{eqnarray}
 \phi_{c,i}(\v{r},t) &=& \phi_{\rm eq} \delta_{1,i} + \delta \phi_{c,i}(\v{r},t)  \\
f_{ij}(\v{p,\v{r},t}) &=& \foi \delta_{ij}+ \delta  f_{ij}(\v{p,\v{r},t}) 
= {\bar f}_{ji} ( -\p, \r , t) \\
g_{ij}(\v{p,\v{r},t}) &=&  \delta  g_{ij}(\v{p,\v{r},t}) = {\bar g}_{ji}^* \prt
\end{eqnarray} 
where we have used the equilibrium conditions: 
$f^{\rm eq.}_{ij}= \foi \delta_{ij}$, $g^{\rm eq.}_{ij}=0$.
We further assume that these deviations are caused by a perturbation 
in the symmetry-breaking external field $\delta h_i \rt$.

Then, by linearizing the non-linear Klein-Gordon equation (\ref{NLKGn}) 
with respect to $\delta \phii \rt$, we obtain
\begin{eqnarray}
(\Box +\mu_1^2) \delta \phi_{c,1} \rt &=& -\frac{\lambda}{6} 
\biggl[
3 \phi_{\rm eq} \delta \langle \tphi_1^2 \rt \rangle 
+ \phi_{\rm eq} \sum_{j=2}^N \delta \langle \tilde{\phi}_j^2 \rt \rangle 
\biggr]  + \delta h_1 \rt
\nonumber \\
\label{LNLKG1} \\
(\Box +\mu_i^2) \delta \phi_{c,i} \rt &=&
 -\frac{\lambda}{3} \phi_{\rm eq} \delta \langle \tilde{\phi}_1 \rt  \tilde{\phi}_i \rt \rangle + \delta h_i
 \qquad (\mbox{for } i \ge 2) \label{LNLKG2}
\end{eqnarray}
where the deviations of the fluctuations of the quantum field $ \tphi_i \rt $ from its equilibrium value
are given in terms of the sum of the four Wigner functions:
\be
\delta \langle \tphi_i \rt \tphi_j \rt \rangle &=& \sum_{\p} \frac{1}{2 \omega_{i,\p}}
 \left[ \delta f_{ij} \prt  +  \delta {\bar f}_{ij} \prt \right. \nonumber \\
& & \hskip 3cm  \left. 
 +  \delta g_{ij} \prt + \delta {\bar g}_{ij} \prt 
\right]
\nonumber \\
& = & \sum_{\p} \frac{1}{2 \omega_{i,\p}}
 \left[ \delta f_{ij} \prt  +  \delta f_{ji} \prt \right. \nonumber \\
& & \hskip 3cm  \left. 
 +  \delta g_{ij} \prt + \delta g_{ji}^* \prt \right]
\label{delta_phi_i_phi_j} 
\ee
where in deriving the last line we have used the following symmetry of the Wigner functions:
\be
\delta {\bar f}_{ij} \prt & = & \delta f_{ji} (-\p, \r, t )  \\
\delta {\bar g}_{ij} \prt & = & \delta g_{ji}^* (\p, \r, t )  
\ee

The space-time dependence of the Wigner functions are determined by the linearized equations of motion which read for $\delta f_{ij}$ and $\delta g_{ij}$:
\be
\delta \dot{f}_{ij}(\v{p},\v{r},t)
  &=& i (\omega_{i,\p} - \omega_{j,\p}) \delta f_{ij} %(\v{p},\v{r},t)
  - \frac{1}{2}\big( \frac{1}{\omega_{i,\p}} + \frac{1}{\omega_{j,\p}} \big)
  \v{p} \cdot (\nabla \delta f_{ij}) \nonumber \\
 & & \quad 
- \left[i \delta U_{ij} %(\v{r},t) %( \foi - \foj)
 + \frac{1}{4}\big( \frac{1}{\omega_{i,\p}^2} - \frac{1}{\omega_{j,\p}^2} \big)
   \v{p} \cdot ( \nabla \delta U_{ij}) \right]  ( \foi - \foj)
      \nonumber \\  & & \qquad + 
  \frac{1}{2} i \nabla \delta U_{ij} \cdot \nabla_{\v{p}} (\foi + \foj) \label{deltafij} \\
 \delta \dot{g}_{ij}(\v{p},\v{r},t)
  &=& -i (\omega_{i,\p} + \omega_{j,\p}) \delta g_{ij} 
  + \frac{1}{2}\big( \frac{1}{\omega_{i,\p}} - \frac{1}{\omega_{j,\p}} \big)
  \v{p} \cdot ( \nabla \delta g_{ij}) \nonumber \\
& & \quad 
- \left[ i\delta U_{ij} 
 + \frac{1}{4}\big( \frac{1}{\omega_{i,\p}^2} - \frac{1}{\omega_{j,\p}^2} \big)
   \v{p} \cdot ( \nabla \delta U_{ij}) \right]  ( \foi + \foj)
      \nonumber \\ & & \qquad
 + \frac{1}{2}  \nabla \delta U_{ij} \cdot \nabla_{\v{p}} ( \foi - \foj) \label{deltagij}
\ee
and the equations of motion for $\delta {\bar f}_{ij}$ and $\delta {\bar g}_{ij}$ 
can be obtained from these by making use of the symmetry properties,  
$ \delta {\bar f}_{ij} (\p, \r, t) = \delta f_{ji} ( - \p, \r, t) $ and 
$ \delta {\bar g}_{ij} (\p, \r, t) = \delta g_{ji}^* ( \p, \r, t ) $.

The deviation of the mean-field potential $\delta U_{ij}$ 
is caused by the change in the change in the condensate and the fluctuation:
\be
\delta U_{ij} (\p, \v{r},t) &=& \frac{\delta \Pi_{ij} (\v{r},t) }{2\sqrt{\omega_{i,\v{p}}\omega_{j,\v{p}}}} 
\label{deltaUij}  
\ee
with
\be
\delta \Pi_{ij}  (\v{r},t) 
&=& \frac{\lambda}{6} \left[ 
2  \phi_{\rm eq} \delta \phi_{c,1} \rt + \sum_{k = 1}^N \delta \langle \tphi_k^2 \rt \rangle 
 \right] \delta_{ij} + \frac{\lambda}{3} \delta \langle \tilde{\phi}_i \rt \tilde{\phi}_j \rt \rangle 
 \nonumber \\
& & \qquad + \frac{\lambda}{3}
 \left(  \phi_{\rm eq.} \delta \phi_{c,i} \rt \delta_{1,j} + \phi_{\rm eq.} \delta \phi_{c,j} \rt \delta_{1,i} \right) ,
\label{deltaPij}
\ee
We solve the linearized equations of motions for a monochromatic perturbation
\be
\delta h_i  \rt  =  e^{-i \omega_+ t + i \v{k} \cdot \v{r}} \delta h_i  + {\rm c. c.} ,
\label{hi}
\ee 
which is adiabatically switched on ($\omega_+ = \omega + i \epsilon$),  
with the following Ansatz:
\begin{eqnarray}
\delta \phi_{c,i} \rt &= &  e^{-i \omega_+ t + i \v{k} \cdot \v{r}} \delta \phi_i + {\rm c. c.} \\
\delta f_{ij} \prt &=&  %\delta {\bar f}_{ji} (-\p, \r, t ) = 
e^{-i \omega_+ t + i \v{k} \cdot \v{r}} \delta f_{ij} ( \bp ) + {\rm c. c.}  
\label{df}\\
\delta g_{ij} \prt &=&  
e^{ - i \omega_+ t + i \v{k} \cdot \v{r}} \delta g_{ij} ( \bp) + {\rm c. c.} \label{dg} 
\end{eqnarray}
where c. c. implies the complex conjugate.
Resulting  linearized kinetic equations are shown in appendix D. 
From these equations, we find two kinds of dispersion relations of collective excitations near equilibrium. One of the relations is essentially equivalent to the dispersion relation of one-component scalar model obtained in I: it gives the dispersion relation of the sigma meson like collective mode. 
The other relation contains pion-like modes which can be interpreted as the Nambu-Goldstone 
modes of $O(N)$ linear sigma model in the chiral limit. 

\subsection{Collective sigma modes in the chiral limit}
As seen in the linearized non-linear  Kein-Gordon equation (\ref{LNLKG1}) 
the fluctuation of the sigma meson condensate $\delta \phi_{c, 1} \rt$ caused by the
external perturbation $\delta h_1 \rt$ couples with the change 
in the diagonal components of the fluctuations $\delta \langle \tphi_i^2 \rt \rangle$.
Along with the Ansatz (\ref{df}) and (\ref{dg}) we put
\be
\delta \langle \tphi_i^2 \rt \rangle = \calf_i e^{-i \omega_+ t + i \v{k} \cdot \v{r}}  + {\rm c. c.}
\label{deltaphi_i2}
\ee
so that
\begin{eqnarray}
\mathcal{F}_i   
&=& \sum_{\p} \frac{1}{2 \omega_{i,\p}}
 \left( \delta f_{ii}  ( \p )   +  \delta f_{ii} ( - \p ) + \delta g_{ii} ( \p ) + \delta g_{ii}^* ( \p ) 
\right) 
 \label{Fi}
\end{eqnarray}
From (\ref{deltafij}) and (\ref{deltagij}) we obtain 
\begin{eqnarray}
\delta f_{ii} ( \p ) &=& 
   \frac{\beta \bv_{i, \p} \cdot \k \left( 1 + \foi \right) \foi }{\omega_+ - \bv_{i, \p} \cdot \k } 
\delta U_{ii} ( \p )  , \label{deltafii} \\
\delta g_{ii} ( \p )  & = & 
\frac{ 2 \foi }{\omega_+ - 2\omega_{i, \p}}
  \delta U_{ii} ( \bp ) , \label{deltagii}
\end{eqnarray}
where $\bv_{i, \p} =  \nabla_{\p} \omega_{i, \p} = \p/\omega_{i,\p} $ and
we have used
\be
\nabla_{\p} \foi  =  - \bv_{i, \p} \beta \left( 1 + \foi \right) \foi 
\ee
while the change of the mean field potential is given by
\be
\delta U_{ii} (\p) &=& \frac{\lambda}{12\omega_{i,\v{p}}} \left[ 
2  \phi_{\rm eq} (2\delta_{1,i} + 1) \delta \phi_1 + 2 \calf_i 
 + \sum_{j = 1}^N \calf_j 
 \right] 
  \label{deltaUii}
 \ee
and the deviation of the sigma meson condensate is given from (\ref{LNLKG1}) 
 \be
 \delta \phi_1  = \frac{1}{\omega_+^2 -k^2 -\mu_1^2} \left[
 \frac{\lambda\phi_{\rm eq}}{6} \left( 3 \calf_1  + \sum_{j= 2}^N \calf_j  \right) +  h_1 \right]
 \label{deltaphi1}
 \ee
These equations form a closed set of equations to determine the linear response of the system 
to the external perturbation $\delta h_1$.   
In finding the solutions it greatly simplifies to observe that the solutions have degeneracy 
due to the symmetry of the equilibrium solutions, 
$\mu_2 = \dots = \mu_N$ which implies $\calf_2 = \dots = \calf_N$ and  
$\sum_{j=2}^N \calf_j = (N-1) \calf_2$. 
We find from (\ref{deltaUii}) and (\ref{deltaphi1}) 
\be
\delta U_{11} & = & \frac{\lambda}{12\omega_{1,\v{p}}} \left[
\left( 1 + \frac{\lambda \phi_{\rm eq}^2}{\omega_+^2 - k^2 - \mu_1^2}  \right)
\left( 3 \calf_1 + (N-1) \calf_2 \right) \right.
\nonumber \\
& & \left. \hskip 6cm + \frac{3 \phi_{\rm eq}h_1}{\omega_+^2 - k^2 - \mu_1^2}  \right]
\label{deltaU11} \\
\delta U_{22} & = & \frac{\lambda}{12\omega_{2,\v{p}}} \left[
\left( 1 + \frac{\lambda \phi_{\rm eq}^2}{\omega_+^2 - k^2 - \mu_1^2}  \right)
\left(  \calf_1 + \frac{1}{3} (N-1) \calf_2 \right) \right.
\nonumber \\
& & \left. \hskip 4cm  +  \frac{2}{3}(N+2) \calf_2 + \frac{ \phi_{\rm eq}h_1}{\omega_+^2 - k^2 - \mu_1^2}  \right]
\label{deltaU22} 
\ee

Inserting (\ref{deltafii}), (\ref{deltagii}) for $i = 1, 2$ in the right hand side
of (\ref{Fi}) and using (\ref{deltaUii}) and (\ref{deltaphi1}) 
we find self-consistency conditions for $\calf_1$ and $\calf_2$:
\begin{eqnarray}
\mathcal{F}_1 &=& - \frac{\lambda}{12} 
  \, \left[ \calk_1 (\omega_+,\k) \left( 3 \mathcal{F}_1 + (N - 1) \mathcal{F}_2 \right) 
  + \calj_1(\omega_+,\k)  \right] \Phi (\omega_+, \bk, \mu_1)
 \label{F1}  \\
\mathcal{F}_2 &=& - \frac{\lambda}{36} 
  \,  \left[ \calk_1 (\omega_+,\k)  \left(  3\mathcal{F}_1 + (N-1) \mathcal{F}_2 \right) 
   + 2 (N + 2) \mathcal{F}_2 + \calj_1 (\omega_+,\k) \right]
\nonumber \\
& & \hskip 5cm \times %\left.   
\Phi (\omega_+, \bk, \mu_2) ,
\label{F2}
\end{eqnarray}
where 
\begin{eqnarray}
\calk_1(\omega,\k) & =  &1 + \frac{ \lambda \phi_{\rm eq.}^2}{\omega^2 - \k^2 - \mu_1^2} 
\label{K1} \\
\calj_1 (\omega,\k)& = & \frac{\phi_{\rm eq.} h_1 }{\omega^2 - \k^2 - \mu_1^2} 
\end{eqnarray}
may be considered as renormalizations of the quasi-particle interaction and 
the coupling to the external field, respectively, due to the coupling to the fluctuation 
of the meson condensate
and $\Phi (\omega_+, \bk, \mu)$ is the function introduced in I:
\begin{equation}
\Phi (\omega_+, \bk, \mu) = \Phi_1 (\omega, \bk, \mu) + i \Phi_2 (\omega, \bk, \mu)
\end{equation}
with the real and the imaginary parts 
\begin{eqnarray}\label{Phi1}
\Phi_1 (\omega, \k, \mu) & = & -
{\cal P} \int  \frac{d^3 \bp}{(2 \pi)^3}  \frac{1}{\wp^2} 
\left[   \frac{ 4 \omega_\bp f_{\rm eq.} ( \wp ) }{ \omega^2 - 4 \omega_\bp^2 } 
 \right.
\nonumber \\
& & \hskip 1cm \left. + \frac{  \beta ( \bv_\bp \cdot \bk)^2 } {\omega^2 - ( \bv_\bp \cdot \bk )^2} 
\left( 1 + f_{\rm eq.} ( \wp ) \right)  f_{\rm eq.} ( \wp )
  \right]  ~,
\\
\Phi_2 (\omega, \k, \mu ) & = & 
 \int  \frac{d^3 \bp}{(2 \pi)^3}  \frac{1}{\omega^2_\bp} \left[  
 2 \pi \left( \delta ( \omega - 2 \omega_\bp )
- \delta ( \omega + 2 \omega_\bp ) \right)  f_{\rm eq.} ( \wp ) \right.
\nonumber  \\
& & \left.
+ \pi \beta  \bv_\bp \cdot \bk \left( \delta ( \omega - \bv_\bp \cdot \bk )
- \delta ( \omega + \bv_\bp \cdot \bk ) \right)  %\nonumber \\
%& & \hskip 3cm   \left.  \times 
\left( 1 + f_{\rm eq.} ( \wp ) \right)  f_{\rm eq.} ( \wp )
  \right]
 \nonumber  \\
 \label{Phi2}
 \end{eqnarray}
 respectively, where we have indicated the quasiparticle mass $\mu$ dependence explicitly.

Eqs. (\ref{F1}) - (\ref{F2}) forms two coupled linear equations of
$\calf_1$ and $\calf_2$ with an inhomogenious terms proportional to $\calj_1$
which may be written in the matrix form:
\begin{equation}
\mathbf{M} \left(
  \begin{array}{c} 
     \mathcal{F}_1 \\
      \mathcal{F}_2 \\
  \end{array}
  \right)
 = 
  \left(
 \begin{array}{c} 
    \mathcal{Q}_1  \\
     \mathcal{Q}_2 \\
  \end{array}
  \right) 
   \calj_1
  \label{trancSC} %trancated Self-Consistency Condition
\end{equation}
with
\begin{equation}
\mathbf{M} =
\left(
  \begin{array}{cc} 
 1 +  \frac{\lambda}{4} \calk_1\Phi (1) 
      &  \frac{\lambda}{12} \calk_1 (N-1) \Phi (1) \\
     \frac{\lambda}{12} \calk_1 \Phi (2)    
     & 1 + \frac{\lambda}{36} \biggl(  \calk_1(N-1) + 2 ( N + 2) \biggr)  \Phi (2)  \\
  \end{array}
\right)
\end{equation}
and
\be
    \mathcal{Q}_1  & = & -  \frac{\lambda}{12}  \Phi (1) \\
     \mathcal{Q}_2 & = &  -  \frac{\lambda}{36}  \Phi (2)
\ee
where we have used abbreviate notations $\Phi (1)  =  \Phi (\omega, \bk , \mu_1)$
and $\Phi (2)  =  \Phi (\omega, \bk , \mu_2)  $.

Solving (\ref{trancSC}) for $\mathcal{F}_1$ and $\mathcal{F}_2$, we find
\be
\mathcal{F}_1 & = &   \frac{ 1 }{\det{\mathbf{M}} } 
 \left(  1+  \frac{\lambda}{18} (N+2) \Phi (2) \right)  \mathcal{Q}_1  \calj_1 
 \label{F_1} \\
\mathcal{F}_2 & = & 
 \frac{1}{\det{\mathbf{M}} }  \mathcal{Q}_2  \calj_1 
 \label{F_2}
\ee
where
\begin{eqnarray}
\det{\mathbf{M}}  =  \left( 1 + \frac{\lambda}{4}  \calk_1 \Phi (1) \right) 
 \left(1 +  \frac{\lambda}{18}   (N+2) \Phi (2)  \right)
\ +\frac{\lambda}{36} \calk_1 (N-1) \Phi (2) 
\nonumber \\
\label{detM}
\end{eqnarray}
These relations determine the linear response of the fluctuations $\delta \langle \tphi_i^2 \rangle$ 
to the external perturbation $h_1$.   
From (\ref{deltaphi1}) 
\begin{eqnarray}
\delta \phi_1 &=& \frac{\lambda}{6} \frac{\phi_{\rm eq.} }{ \omega_+^2 - k^2 - \mu_1^2 } 
 \left( 3 \mathcal{F}_1 +  (N-1) \mathcal{F}_2 \right)  
+ \frac{1}{ \omega_+^2 - k^2 - \mu_1^2 }  \delta h_1
\end{eqnarray}
which upon insertion of (\ref{F_1}) and (\ref{F_2}) gives
\be
\delta \phi_1 = \frac{1- \mathcal{S}_1 }{ \omega_+^2 - k^2 - \mu_1^2 } \delta h_1
\ee
with
\begin{eqnarray}
\mathcal{S}_1 &=& 
\frac{\lambda \phi_{\rm eq.}
\left[   \Phi (1) \left(  1+  \frac{\lambda}{18} (N+2) \Phi (2) \right)  + \frac{N- 1}{9} \Phi (2) \right]
}{ 24 ( \omega_+^2 - k^2 - \mu_1^2) \det{\mathbf{M}} } .
\end{eqnarray}
The quantity $\mathcal{S}_1$ represents the medium modification of the propagation of the
sigma meson field due to the absorption and the reemission of the fluctuation by the pair of
quasi-particle excitations.
We note that $\mathcal{S}_1$ is regular on the quasi-particle mass-shell 
$\omega^2 = k^2 + \mu_1^2$ while new singularity appears when the condition 
\be
 \det{\mathbf{M}} (\omega, \k ) = 0
\label{detM=0}
\ee
is satisfied, corresponding to the collective excitations of the system.  

We have shown in I that the undamped collective mode appears in the time-like region 
in the low temperature phase where there exists non-vanishing sigma meson condensate
$(\phi_{\rm eq.} \ne 0)$.   
To eliminate the singularity of $\mathcal{K}_1$ at meson poke we multiply 
(\ref{detM=0}) by $ \omega^2 - k^2 - \mu_1^2$.  One then finds at $k =0$
\be
( \omega^2 - \mu_1^2 ) & &  \left(1 +  \frac{\lambda}{18}   (N+2) \Phi (2)  \right) 
 + \frac{\lambda}{4} \left( \omega^2 - \mu_1^2 + \lambda \phi_{\rm eq.}^2 \right)  
 \nonumber \\ 
& & 
\times  \left[ \Phi(1)  \left(1 +  \frac{\lambda}{18}   (N+2) \Phi (2)  \right) + \frac{N-1}{9} \Phi (2)  \right]
= 0
\ee
for the condition of the shifted meson pole. 

\begin{figure}[h]
\begin{center}
\includegraphics[scale=1]{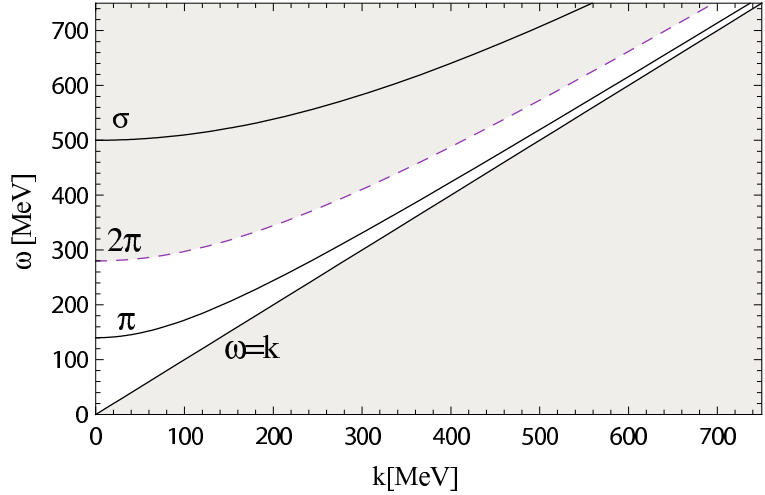}%Kin_sigma.eps}
\end{center}
\caption{The region where the imaginary part of $\Phi ( \w + i \epsilon, k $) has non-vanishing
value}
\label{fig3}
\end{figure}

The entire space-like region $\omega < k$ is covered by the quasiparticle continuum 
associated with the scattering of the thermally populated quasiparticle excitations which 
gives the non-vanishing imaginary parts when the kinematical condition 
$\omega = \epsilon_{i, \p + \k} - \epsilon_{i,\p} \simeq \k \cdot \bv_{i,\p}$ is fulfilled.  
Since $v_{i,\p} = p /\omega_{i,\p}  \simeq 1 - \mu_i/ p$, this condition is met near
the light cone $\omega = k$ only for quasiparticles with large values of $p$ or 
$\omega_{i,\p}$ whose number is suppressed by the Boltzmann factor $\foi$
Indeed, we have found in I that the integral over $\p$ in (\ref{Phi2}) can be carried 
out analytically, yielding 
\be
\Phi_2 (\omega, \k, \mu ) & = & 
\frac{1}{8\pi^2} 
\frac{\omega}{k} \frac{1}{e^{\frac{\mu\beta}{\sqrt{1 - (\omega/k)^2}}} -1}  \theta ( k - \omega) 
\nonumber \\
& & \hskip 3cm 
+ \frac{\sqrt{\omega^2 - 4 \mu^2}}{2 \pi \omega } \frac{1}{e^{\omega \beta/2} - 1}
 \theta (\omega - 2\mu)
\nonumber \\
\label{Phi20}
\ee
The first term gives continuum in the whole space-like region which diminishes
exponentially as one approaches to the edge $\omega = k$ due to the Bose-distribution
factor $\foi = (e^{\beta\omega_\p} - 1)^{-1}$ for 
$\omega_\p = (\omega/k) p = \mu/ \sqrt{1 - (\omega/k)^2}$. 

The second term in (\ref{Phi20}) may be interpreted as due to the {\it thermally  induced} 
pair creation or annihilation, namely one of the pair of quasiparticles created or annihilated
has been present as thermal excitations.
It gives continuum in the time-like region 
$\omega = \omega_{\p +\k} + \omega_\p \simeq 2 \omega_\p \ge 2\mu$.  
This continuum thus covers the region for the $\omega$ greater than the twice the lightest 
quasiparticle mass, namely $\mu_2$ 
Since the sigma meson mass $\mu_1$ may be greater than two times the pion-like 
quasiparticle mass $\mu_2$, the collective mode of the sigma meson like mode
has additional thermal decay width in addition to the pair decay width in the vacuum.

For a long wavelength, low energy excitation mode ($\w, k \ll \mu_i$), $\calk_1(\w,\k)$ may be approximated as
\begin{eqnarray}
\calk_1(\omega,\k) \simeq \calk_1 (0,0) =1 - \frac{\lambda \phi_{\rm eq.}^2}{\mu_1^2} 
\end{eqnarray}
In the exact chiral limit $(\ep=0)$, 
\be
\lambda \phi_{\rm eq.}^2 = \
\begin{cases}
 - 3 \mu_1^2 &(\phi_{\rm eq.} \neq 0)\\
 0 &(\phi_{\rm eq.} = 0)
\end{cases}
\ee
so that we have 
\be
\calk_1 (0,0) =
\begin{cases}
 -2 &(\phi_{\rm eq.} \neq 0)\\
 1 &(\phi_{\rm eq.} = 0)
\end{cases}
\ee
which gives %In this case the dispersion relation is determined by 
\begin{eqnarray}
\det{\mathbf{M}} = 1  - \frac{\lambda}{2} \Phi (1) +  \frac{\lambda}{6} \Phi (2) 
- \frac{\lambda^2}{36} (N+2) \Phi (1) \Phi (2) 
\label{O2disp_low}
\end{eqnarray}
for low temperature symmetry broken phase ($\phi_{\rm eq.} \neq 0$)  and 
\begin{eqnarray}
\det{\mathbf{M}} = \left( 1 +  \frac{ \lambda}{12} ( N+2 ) \Phi (1) \right) 
\left( 1 -  \frac{\lambda}{6}  \Phi (1)  \right) 
\label{O2disp_high}
\end{eqnarray}
for high temperature symmetric phase ($\phi_{\rm eq.} = 0$) where $\Phi (1) = \Phi (2)$.

We show in Fig. 4 the numerical result of the shifted pole mass $\mu_1'$ of the collective
sigma-like excitation in the long wavelength limit ($k = 0$).   
At zero temperature $\mu_1'$ coincides with $\mu_1$.   
As the temperature increases the coupling to the thermally excited quasiparticles makes
the pole mass $\mu_1'$ smaller than $\mu_1$.
The mass shift exhibits non-analytic kink behaviors as it crosses the quasiparticle threshold 
at $\omega = 2 \mu_2$ and at certain temperature $T_1$ slightly above $T_c$ it 
becomes zero indicating the spinodal instability as discussed in I.  
This instability coincides with the condition for the appearance of the 
aforementioned tachyonic mode.

\begin{figure}[h]
\begin{center}
\includegraphics[scale=1]{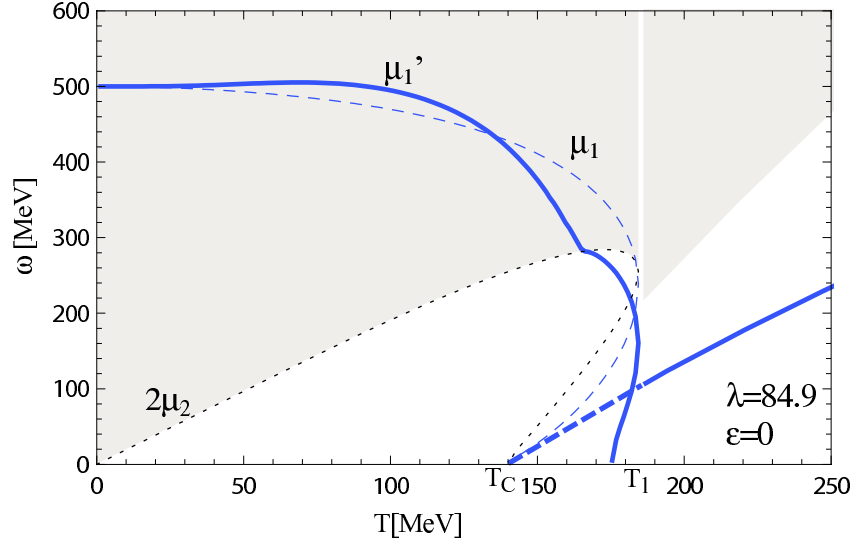}%Collective_Sigma_CL.eps}
\end{center}
\caption{The spectrum of sigma-like excitations with $k=0$.  The effective mass 
$\mu'_1$ of the collective sigma mode in the chiral limit 
($\epsilon =0$) is plotted by thin solid curve.  The continuum of the sigma-like 
quasi-particle excitations are shown by the gray areas for low temperature phase
($T < T_1$) and for high temperature phase ($T < T_1$) where $T_1$ is highest 
temperature where the gap equation has non-zero solution for $\phi_{\rm eq.}$. }
\label{fig4}
\end{figure}

%In the case of $\ep \neq 0$, ...

\subsection{Collective pionic modes in the chiral limit: the Nambu-Goldstone modes}

We now compute the collective excitations carrying the pionic quantum number which 
couple to the oscillations of the meson condensate in the direction perpendicular 
to the equilibrium condensate, namely $\delta \phi_i$.
First we show that the missing Nambu-Goldstone modes appear as collective excitations 
with phonon-like dispersion relation in the exact chiral limit.
 
The equations of motion of $\delta \phi_i$ ,  (\ref{LNLKG2}), indicate
that pionic modes couple to the fluctuation of the type $\delta \langle \phi_1 \phi_i \rangle$.
There are $N-1$ such mode for $i = 2, \dots, N$, but these modes are all degenerate
due to the unbroken $O(N-1)$ symmetry, corresponding to the isospin symmetry in case of
$N=4$.
So it suffices to compute the $i=2$ component to obtain the pionic excitation spectrum.   

Applying a monochromatic perturbation with non-vanishing $\delta \phi_2$, we set 
\be
\delta \langle \tphi_1 \tphi_2 \rangle \rt = \calf_{12} e^{-i \omega_+ t + i \k \cdot \r} + {\rm c. c.}
\ee
with
\begin{eqnarray}
\mathcal{F}_{12} 
=  
 \sum_{\p} \frac{1}{2\sqrt{ \omega_{1,\p} \omega_{2,\p}}}
\left( \delta f_{12} (\p)  +  \delta {\bar f}_{12} (\p) 
+ \delta g_{12} (\p) + \delta {\bar g}_{12} (\p) 
\right) 
\label{calf12}
\end{eqnarray} 
From the linearized equations of motion of the corresponding Fourier components 
of the Wigner functions given in Appendix D, we obtain
\be
\delta f_{12} (\p)  & = &
- \calg_{12}^{(-)} (  \omega_+, \k, \p ) \caln_{12}^{(-)} ( \p)
\delta U_{12}
 %\nonumber 
 \\
% & & \left(  \omega_+ + (\wa - \wb) + (\p \cdot \k /2) (1/\wa + 1/\wb) \right)
 \delta {\bar f}_{12} (\p) 
& = &  - \calg_{12}^{(-)}  ( - \omega_+, \k, \p ) \caln_{12}^{(-)} ( \p)
\delta U_{12}
 %\nonumber 
 \\
\delta g_{12} (\p) & = &
- \calg_{12}^{(+)} ( - \omega_+, \k, \p ) \caln_{12}^{(+)} ( \p)
\delta U_{12}
 %\nonumber 
 \\
\delta {\bar g}_{12} (\p) & = &
- \calg_{12}^{(+)}  (  \omega_+, \k, \p ) \caln_{12}^{(+)} ( \p)
\delta U_{12}
\ee
where
\be
 \calg_{12}^{(\mp)}  (\omega, \k ; \p ) & = & 
\frac{1}{  \omega - (\wa \mp \wb) - \k \cdot ( \bv_{1, \p} \pm  \bv_{2,\p} ) / 2} \\
\caln_{12}^{(\mp)} ( \k, \p ) & = &\left( 1 +  \frac{\k \cdot \p}{4} \left( \frac{1}{\wa^2} - \frac{1}{\wb^2} \right) \right)
 (\foa \mp \fob) \nonumber \\
 & & \quad \hskip 3cm + (\k/2)\cdot \nabla_{\p} (\foa \pm \fob)
 \nonumber \\
\ee
are the propagators of a pair of quasiparticles of two different types and the effective 
number of such states, respectively. 
The off-diagonal components of the disturvance caused in the mean-field potential may be 
obtained from (\ref{Uij}) 
\be
\delta U_{12} (\p)  =  \frac{\delta \Pi_{12}}{2\sqrt{\omega_{1,\p} \omega_{2,\p}}} 
\ee
with 
\be
\delta \Pi_{12} =  \frac{\lambda}{3} \left( \phi_{\rm eq.} \delta \phi_2 + \calf_{12} \right)
\ee
Using the solution of the linearized equation of motion for a classical condensate (\ref{LNLKG2}) 
%we have 
\begin{eqnarray}
\delta \phi_2
= \frac{1}{\omega_+^2 - k^2 - \mu_2^2} 
\left[ \frac{\lambda}{3} \phi_{\rm eq.} \calf_{12} -  \delta h_2 \right]
\end{eqnarray}
we find
\begin{eqnarray}
\delta U_{12} = \frac{\lambda}{6} 
 \frac{1}{\sqrt{\omega_{1,\p}\omega_{2,\p}}} 
 \left( \calk_2 (\omega_+, \k)  \calf_{12} - \calj_2 (\omega_+, \k) \right)
 \label{du12}
\end{eqnarray}
where
\be
\calk_2 (\omega,\k) & =  &
1 + \frac{\lambda}{3} \, \frac{\phi_{\rm eq.}^2}{\omega^2 - \k^2 - \mu_2^2}  
\label{K2} \\
\calj_2  (\omega,\k) & = & \frac{1}{\omega^2 - k^2 - \mu_2^2} \phi_{\rm eq.}  \delta h_2
\label{h2}
\ee
Inserting (\ref{du12}) into (\ref{calf12}) we find the self-consistency condition for
the pionic excitations:
\begin{eqnarray}
\calf_{12} = \frac{\lambda}{12} \Psi (\omega_+,\k) 
\left[ \calk_2 (\omega_+,\k) \calf_{12} - \calj_2 (\omega_+, \k) \right]
\label{calf12}
\end{eqnarray}
where 
\begin{eqnarray}\label{Psi}
\Psi (\omega,\k) 
& = & \sum_{\p} \frac{1}{\omega_{1,\p}\omega_{2,\p}} A (\w, \k, \p) 
\label{Psi}
\ee
with
\be
A (\w, \k, \p) 
 %\nonumber \\
 & = & 
% \left[
 \left( \calg_{12}^{(-)}  ( -\w, \k, \p) + \calg_{12}^{(-)}  ( \w, \k, \p) \right) \caln_{12}^{(-)} ( \k, \p ) 
 %\right.
\nonumber \\
& & \qquad %\left.
+ \left( \calg_{12}^{(+)}  ( - \w, \k, \p) + \calg_{12}^{(+)}  ( \w, \k, \p) \right) \caln_{12}^{(+)} ( \k, \p ) 
%\right]
\end{eqnarray}
is a modification of  the dimensionless function $\Phi (\omega, \k)$ which has appeared 
in the computation of the sigma meson collective modes.  Indeed, if we set 
$\mu_1 = \mu_2 = \mu$, the function $\Psi (\omega , \k )$ coincides with $\Phi (\omega, \k )$.

\begin{figure}[h]
\begin{center}
\includegraphics[scale=1]{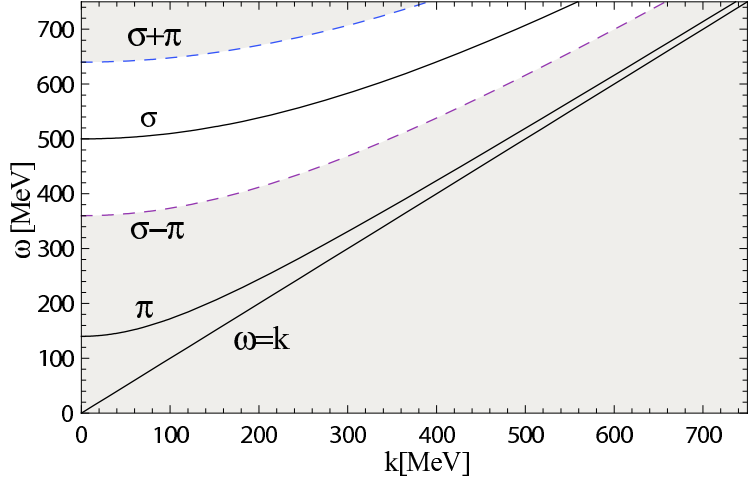}%kin_pi.eps}
\end{center}
\caption{The region where the imaginary part of $\Psi ( \w + i \epsilon, k $) has non-vanishing
value is shown by gray area.}
\label{fig5}
\end{figure}

Solving (\ref{calf12}) for $\calf_{12}$ we find the expression for the linear response of the
fluctuations to the exterrnal perturbation $\calj_2$:
\be
\calf_{12} = \frac{\frac{\lambda}{12} \Psi (\omega_+,\k) }{1 - \frac{\lambda}{12} \Psi (\omega_+,\k)  \calk_2 (\omega_+,\k) } \calj_2
\ee
The condition that Eq. (\ref{calf12}) contains non-trivial solution gives the dispersion relation
\begin{eqnarray}
\frac{\lambda}{12} \mathcal{K}_2 (\omega,\k) \Psi (\omega,\k) = 1  
\label{D12}
\end{eqnarray}

Now we show that this equation has the solution of the form 
\begin{equation}\label{phonon}
\omega = \bv \cdot \bk
\end{equation}
which may be identified as the missing Nambu-Goldstone modes. 

\begin{figure}[b]
\begin{center}
\includegraphics[scale=1]{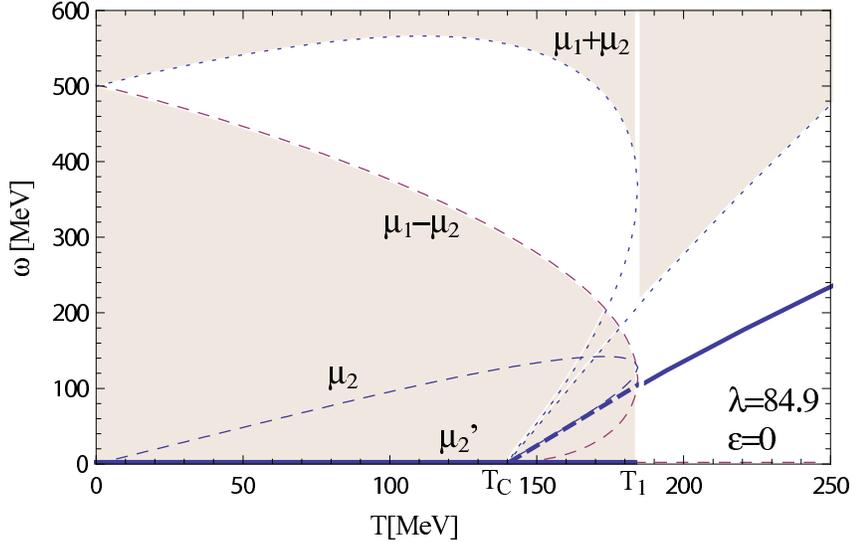}%Collective_Pi_CL.eps}
\end{center}
\caption{The spectrum of pion-like excitations with $k=0$.  The Nambu-Goldstone 
mode in the low temperature broken-symmetry phase at $T < T_1$ is shown by 
the thick solid line as well as the massive pion mode 
degenerate with sigma-meson mode in the high temperature phase.
The continuum of the sigma-like quasi-particle excitations are shown by the 
gray areas for low temperature phase ($T < T_1$) and for high temperature 
phase ($T < T_1$). }
\label{fig6}
\end{figure}

For this purpose, we first show that the dispersion relation (\ref{D12}) 
is satisfied for $(\w, |\k| )=(0,0)$.  
We recall that In the chiral limit ($\epsilon =0$) the static non-linear Klein-Gordon equation 
(\ref{staticNLKG}) and the definition of the quasiparticle masses (\ref{mu_i}) read
\be
m^2 &=& -\frac{\lambda}{6} \left( \phi_{\rm eq.}^2 + 3 \fla_{\rm eq.} + (N-1) \flb_{\rm eq.}  \right) 
\label{m20} \\
\mu_1^2  &=& m^2 +\frac{\lambda}{6}\left( 3 \phi_{\rm eq.}^2 + 3 \fla_{\rm eq.} + (N-1) \flb_{\rm eq.}  \right) 
\label{mu_10} \\
\mu_2^2  &=& m^2 +\frac{\lambda}{6}\left( \phi_{\rm eq.}^2 +  \fla_{\rm eq.} + (N+1) \flb_{\rm eq.}  \right) 
\label{mu_20}  
\ee
respectively.
Inserting (\ref{m20}) into (\ref{mu_20}), we find 
\be
\mu_2^2 = \frac{\lambda}{3} \left(  \flb_{\rm eq.}  - \fla_{\rm eq.} \right) \label{mu22ab}
\ee
From (\ref{mu_10}) and (\ref{mu_10}),  we also find 
\be
\mu_1^2-\mu_2^2
 = \frac{\lambda}{3}\left(  \phi_{\rm eq.}^2 +  \fla_{\rm eq.} - \flb_{\rm eq.}  \right) \label{Difmu1mu2}
 \label{mu1-mu2}
\ee
Then, 
\be
\mathcal{K}_2 (0,0) &=& 
	 1 - \frac{\lambda}{3} \frac{\phi_{\rm eq.}^2}{\mu_2^2} 
	=  \frac{ \phi_{\rm eq.}^2 +  \fla_{\rm eq.} - \flb_{\rm eq.} }{\fla_{\rm eq.} - \flb_{\rm eq.}}
 \label{K200}
\ee
where we have used (\ref{mu1-mu2}).
On the other hand, 
\be 
A (0, 0, \p ) & = &
 \left( \calg_{12}^{(-)}  ( 0, 0, \p) + \calg_{12}^{(-)}  ( 0, 0, \p) \right) \caln_{12}^{(-)} ( 0, \p ) 
\nonumber \\
& & \qquad 
+ \left( \calg_{12}^{(+)}  ( 0, 0, \p) + \calg_{12}^{(+)}  ( 0, 0, \p) \right) \caln_{12}^{(+)} ( 0, \p ) 
\nonumber \\
& = &  2\left( \frac{\foa - \fob}{\w_{1,\p}-\w_{2,\p}} -\frac{\foa + \fob}{\w_{1,\p} + \w_{2,\p}} \right) 
\nonumber \\
& = & \frac{4(\w_{2,\p} \foa - \w_{1,\p} \fob)}{\mu_1^2 - \mu_2^2} 
\label{A00}
\ee
where we have used $\calg_{12}^{(\mp)}  ( 0, 0, \p) = (\w_{1,\p} \mp \w_{2,\p} )^{-1}$,
$\caln_{12}^{(\mp)} ( 0, \p ) = \foa \mp \fob$ 
and $\w_{1,\p}^2-\w_{2,\p}^2= \mu_1^2 - \mu_2^2$.
Inserting (\ref{A00}) into (\ref{Psi}) we obtain 
\be
\Psi (0,0) 
&=&  \frac{4}{(\mu_1^2 - \mu_2^2)} \sum_{\p} \left( \frac{\foa}{\w_{1,\p}} - \frac{\fob}{\w_{2,\p}} \right) 
= \frac{4}{(\mu_1^2 - \mu_2^2)} \left( \fla_{\rm eq.} - \flb_{\rm eq.} \right) \nonumber \\
&=& \frac{12}{\lambda} \cdot
\frac{\fla_{\rm eq.} - \flb_{\rm eq.}}{ \phi_{\rm eq.}^2 +  \fla_{\rm eq.} - \flb_{\rm eq.} } 
\label{Phi1200}
\ee
where we have used (\ref{mu1-mu2}) again.
Comparing (\ref{K200}) and (\ref{Phi1200}) we finally find % the relation:
\be
\frac{\lambda}{12}  \Psi (0,0) 
=  \calk_2 (0, 0 )^{-1}
\label{NGboson}
\ee
which implies that the equation ($\ref{D12}$) has the solution $(\w, \k)=(0,0)$.  

We note that the acoustic Nambu-Goldstone modes appear in the continuum of of the 
quasi-particle excitations where the function $\Psi (\omega_+,\k)$ gives non-vanishing 
imaginary part (see Fig. 5)  thus always suffers the Landau-damping at finite temperature.

%%%%%%%%%% subsection 5-3
\subsection{Collective modes with explicit symmetry breaking $(\ep \neq 0)$}
 
As we have seen in section 4, in the presence of the explicit symmetry breaking $\epsilon \ne 0$, 
the mass parameter $\mu_i$ becomes non-vanishing at zero temperature and
we have interpreted it as the physical pion mass in case of $N = 4$.   
At finite temperatures, the pionic collective modes, which form the massless Nambu-Goldstone
modes in the chiral limit, thus acquire non-vanishing masses.
These modes would also persist in the high temperature region since the transition becomes 
smoothed out as we have seen in the previous section. 
However, there may be some qualitative change in the character of the pionic collective 
modes in the low temperature region and in the high temperature region since the NG bosons 
appear in the chiral limit only in the low temperature phase.  
We now examine these problems. 

\subsubsection{Collective sigma modes with symmetry breaking}
As shown in section 5.1, we have computed collective sigma modes with explicit summetry breaking 
($ \epsilon \neq 0 $). We show in Fig. 7 the numerical result of the shifted pole mass $\mu_1'$ of the 
collective sigma-like modes in the long wavelength limit. At zero temperature $\mu_1'$ is equal to the quasi-particle mass $\mu_1$.  As the temperature increased, $\mu_1'$ slightly exceeds $\mu_1$ at $T<120$MeV, while it becomes smaller than $\mu_1$ at $T>120$MeV.  The sigma mass shift has 
non-analysic kink similar to that in the case of $\epsilon=0$ when it crosses the quasiparticle 
threshold at $\omega = 2 \mu_2$. However, it exhibits no instability since the first order transition 
is smoothed out.   The shifted sigma pole mass also increases above $T=230$MeV where 
the chiral symmetry is restored.

\begin{figure}[h]
\begin{center}
\includegraphics[scale=1]{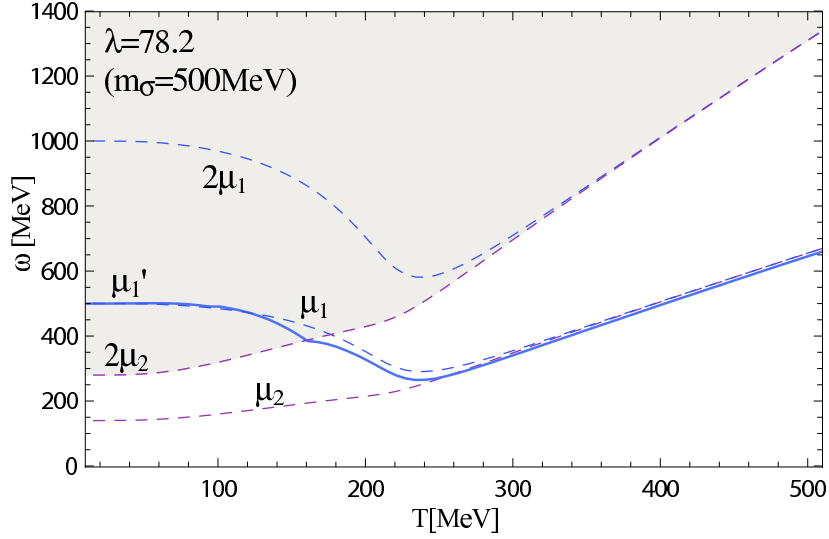}%Collective_sigma_SSB.eps}
\end{center}
\caption{The spectrum of sigma-like excitations with explicit symmetry breaking. 
Effective mass $\mu'_1$ of collective sigma meson mode with explicit symmetry breaking
is shown by the solid line.  The gray area indicates the continuum of quasi-particle 
excitations with the same quantum number as a sigma meson.}
\label{fig7}
\end{figure}

\subsubsection{Collective pionic modes with symmetry breaking}
We have computed the mass shift of the collective pionic excitations with explicit symmetry breaking. 
By using Eq. (\ref{D12}), we show in Fig. 8 the numerical results of the pionic pole mass $\mu_2'$ 
in the long wavelength limit.   At zero temperature $\mu'_2$ is equal to pionic quasi-particle 
mass $\mu_2$.   As temperature increses $\mu'_2$ becomes smaller than $\mu_2$. 
At $150$MeV$<T<170$MeV, it exhibits non-trivial behavior.
It has non-analytic kink where it crosses the quasiparticle threshold $\mu_1 - \mu_2$.

\begin{figure}[h]
\begin{center}
\includegraphics[scale=1]{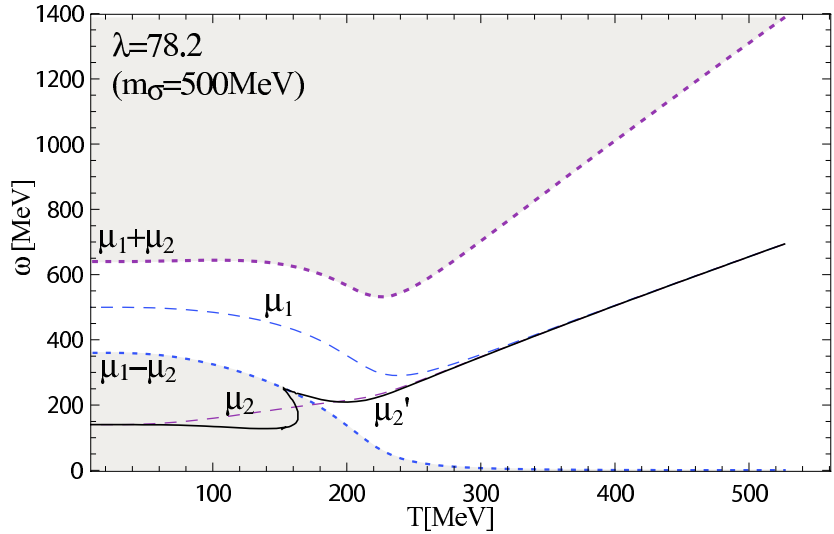}%Collective_pi_SSB.eps}
\end{center}
\caption{The spectrum of pion-like excitations with symmetry breaking.
Effective mass $\mu'_2$ of collective pion mode with explicit symmetry breaking
is shown by the solid line.  The gray area indicates the continuum of quasi-particle 
excitations with the same quantum number as a pion.}
\label{fig8}
\end{figure}

Recently, the shifted pole masses of the sigma/pionic collective modes has been investegated 
also by Tsue and Matsuda\cite{TsueMatsuda2008} based on the earlier work\cite{TVM00} 
in the functional Schr\"odinger picture\cite{KV88,EJP88,VM97}.  
These authors discuss the cut-off dependence of the mass shift which arises due to the 
inclusion of the vacuum polarization, the effect which we have neglected in this paper.  
The kink structure of the mass shift is smoothed out in their results, however. 
This may be due to the static approximation ($\omega=0$) they have taken in evaluating 
the pion self-energy which misses the threshold behavior of the quasi-particle excitations.

\begin{figure}%[H]
\begin{center}
\includegraphics[scale=1]{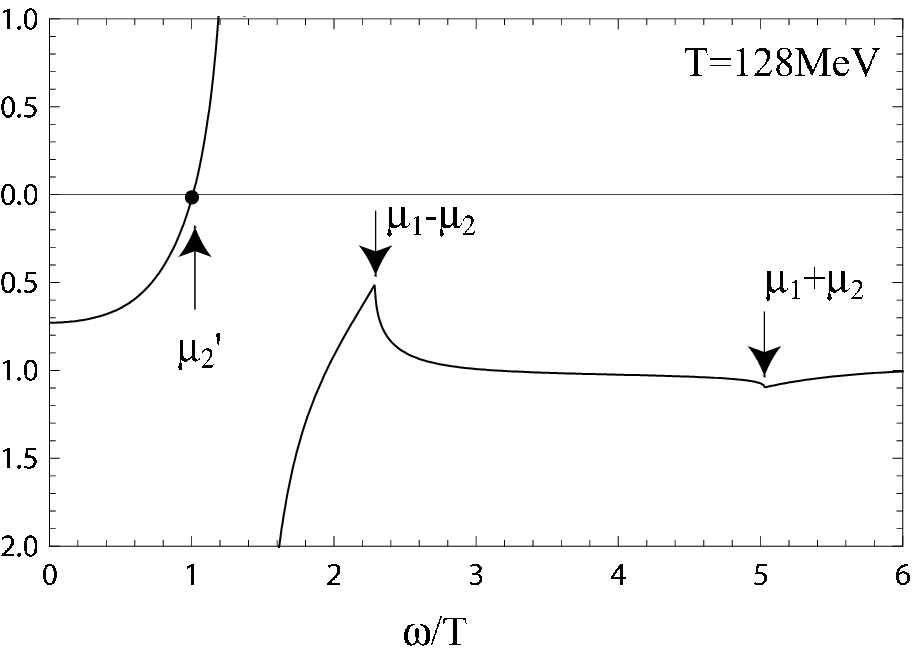}%Real_pi_SSB_a.eps}
\vskip 2pt
\includegraphics[scale=1]{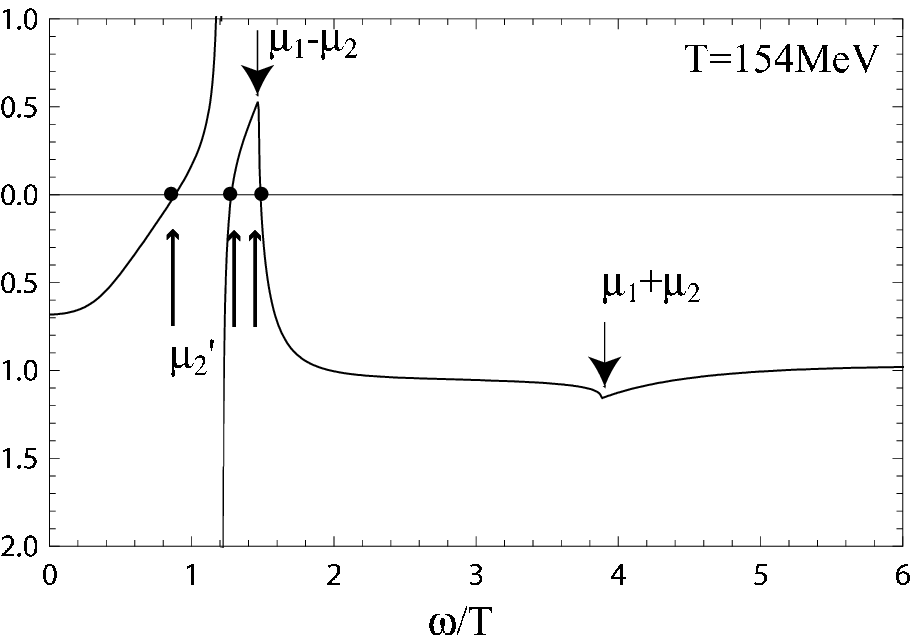}%Real_pi_SSB_b.eps}
\vskip 2pt
\includegraphics[scale=1]{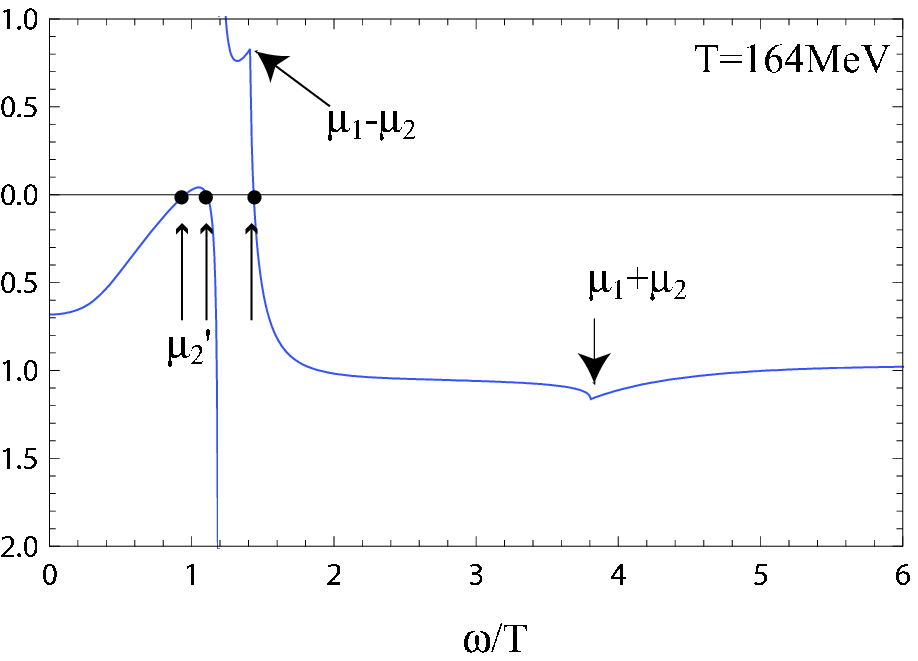}%Real_pi_SSB_c.eps}
\end{center}
\caption{Graphical determination of the effective pion mass  for the case of explicit symmetry 
breaking.  The function defined by Eq. (5.80) is plotted against $\omega$ for $k=0$ at three 
different temperatures as indicated in each frame.  The crossing points with the horizontal axis 
give the effectve masses of collective pionic excitations.}
\label{fig9}
\end{figure}

To clarify this point, we examine the dispersion relations of the long wavelength pionic 
excitation modes for $\epsilon \neq 0$, plotting in Fig.9
\be
D_{\pi}(\omega.T)=\lim_{\k \to 0} \mbox{Re}\left[ \frac{\lambda}{12} \mathcal{K}_2(\omega_{+},\k) \Psi (\omega_{+},\k)-1 \right].
\ee
%epsilon \neq 0$ 
as the function of the excitation energy $\omega$ scaled by the temperature.   
We also plot in Fig. 10 the real part of the function $\Psi (\omega, k)$ which are used to
evaluate $D_{\pi}$ as well as the imaginary part which determines the region of the 
quasi-particle continuum.
The value of $\omega$ at the crossing points of this curve with the horizontal axis give the 
energies of the collective excitations. 
At $T=128$MeV the graph of $D_{\pi}$ has two kinks corresponding to the quasi-particle threshold 
$\mu_1 \pm \mu_2$.   As temperature increases the kinks shift to the left and the kink at $\mu_1-\mu_2$ exceeds the horizontal line.  At $T=160$MeV, therefore, we have three crossing points which corresponds to the non-trivial behavior of the pionic shifted mass as shown in Fig. 7 at $150$MeV$<T<170$MeV.   At low temperatures, the tip of the kink is located below the horizontal axis.  
As the temperature exceeds, the tip of the kink moves upward and hit the horizontal axis 
giving two extra solutions of $D_{\pi}(\omega.T) = 0$.  This is the reason why a pair of 
solutions appear with kink in the plot of the pion masses at the threshold of quasi-particle 
excitations at $\omega = \mu_1 - \mu_2$. 

\begin{figure}%[H]
\begin{center}
\includegraphics[scale=1]{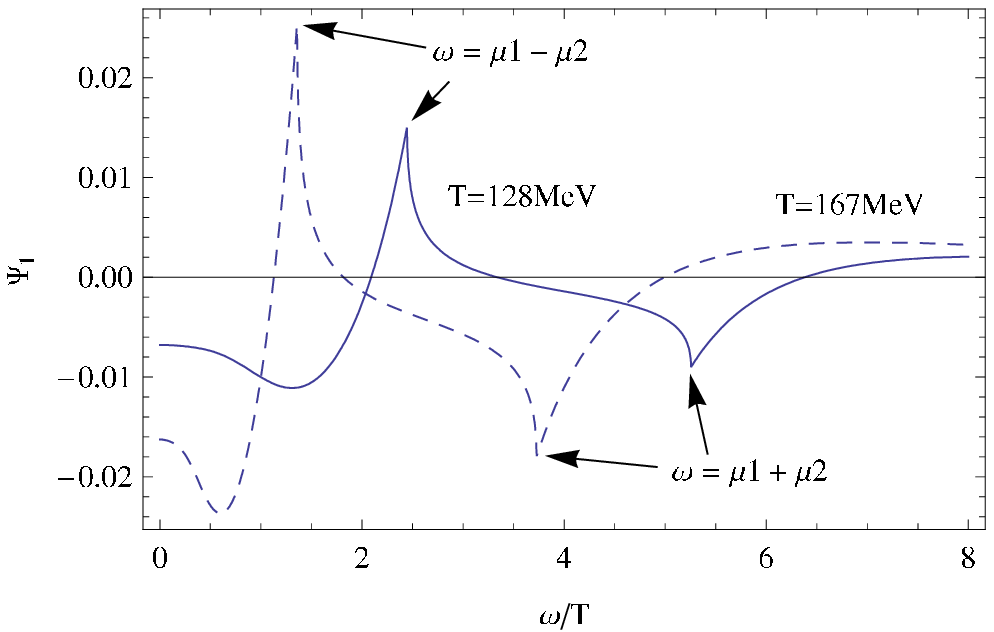}%Psi1_SSB.eps}
\vskip 2pt
\includegraphics[scale=1]{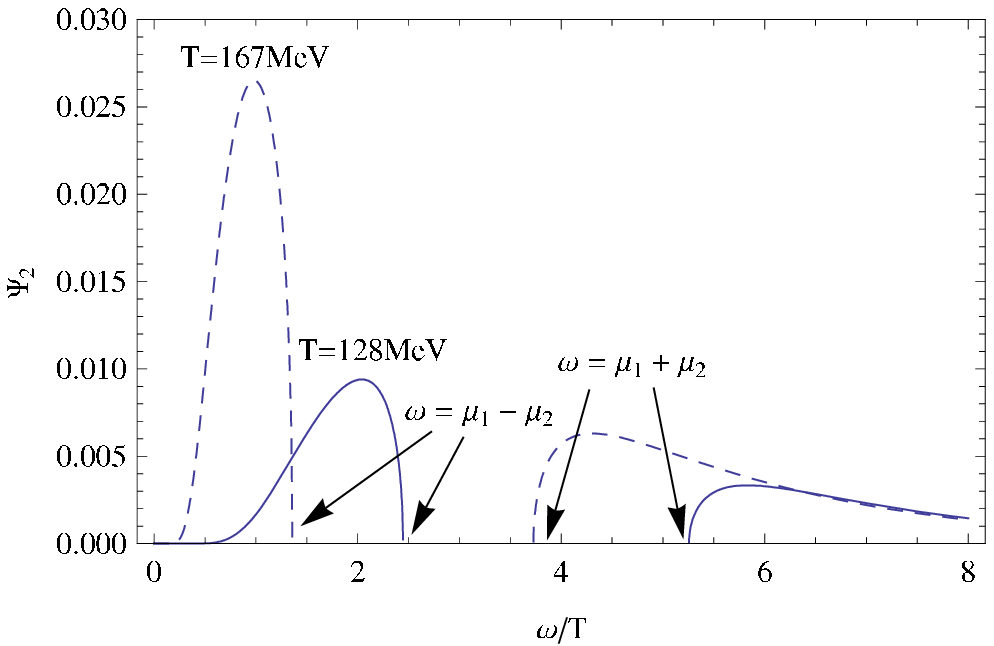}%Psi2_SSB.eps}
\end{center}
\caption{Real part (upper panel) and the imaginary part (lower panel) of the function 
$\Psi(\omega, k)$  at $k=0$}
\label{fig10}
\end{figure}

We note that in the chiral limit the branch which originate from $\mu_2$ at $T=0$ becomes
the massless Nambu-Goldstone modes.    
Our results show that with the explicit chiral symmetry breaking, the Nambu-Goldstone 
mode becomes a massive Landau-damping mode at low temperature which terminates 
at the kink position and there appears a new mode which becomes the pionic  
mode degenerate with the sigma meson mode in the high temperature symmetric phase.

\section{Concluding remarks}
In this paper we have developed a quantum kinetic theory for the chiral condensate and quantum meson excitations based on the formalism we have developed in our previous paper\cite{MM08}
applying to the $O(N)$ sigma model.
We have shown that the time-evolution of system is described by a coupled form of kinetic equations consisting of the $N$ classical non-linear Klein-Gordon field equations for classical meson fields (condensates) and $2N \times 2N$ kinetic equations for the generalized Wigner functions of 
$N$ quantized meson fields.   

We then applied our kinetic theory to describe uniform equilibrium state and to calculate
the dispersion relations of collective modes near equilibrium.     
We constructed time-independent solutions of coupled equation for $O(N)$ model assuming 
only one component of the meson fields has non-vanishing expectation value in equilibrium. 
We recover the results in the Hartree approximation at finite temperature.
The order of the phase transition, calculated from the gap equations, becomes first order in the
chiral limit, while it is smoothed out when we introduce symmetry breaking in order to generate
actual pion mass.
It is also known that the Goldstone theorem (in the chiral limit) is apparently violated because all 
mass parameters in quantized field fluctuations become non-vanishing even in the low temperature 
symmetry broken phase. 

It was shown that the dispersion relation of $O(N)$ model decouples into two types of the relations; 
the dispersion relation of the sigma-like mode and that of pion-like modes.
From the dispersion relation of the sigma-like mode, a fluctuation in the direction of the condensate, 
we obtain only meson-like excitations with mass gap, but no collective phonon mode. 
The effect of the anomalous Wigner functions, $g \sim \langle a a \rangle$, plays an important role 
for preventing the solution of the dispersion relation from the violation of causality as in the case of 
one-component theory.
On the other hand, the dispersion relation of the pion-like modes, for fluctuations in the direction perpendicular to the condensate, contains massless collective mode which can be interpreted as a missing Nambu-Goldstone mode. 

We examined how these mesonic collective modes changes their characters in the absence of the 
exact chiral symmetry.  
It was shown that the Nambu-Goldstone mode becomes massive 
%suffers the Landau-damping at finite temperature, 
and is non-analytically transformed through kink to the pion mode in the symmetric phase. 
Such a kink always appears at the threshold of continuum of the underlying quasi-particle excitations.  
 
Finally, we like to make a few comments on the improvement of our formalism and its applications to 
more realistic physical situations.  We have derived quantum kinetic equations without collision terms using Gaussian Ansatz for the initial density matrix. 
This approximation corresponds to a neglect of all correlations in the system.
Justification of this approximation remains an important problem for further study. 
Recently, a modified self-consistent Hartree approximation in
finite temperature scalar field theory has been  proposed in the
context of chiral phase trantsion\cite{IvanovRiekKnoll2005} and 
Bose-Einstin condensation\cite{Kita2006}. 
Although quasi-particle continuum may be modified by these more sophisticated version of 
the Hartree approximation we expect some features may remain unchanged by such improvements.  
The collision terms is needed to describe relaxation phenomena accompanied by entropy 
production\cite{CalzettaHu1988,Berges2004,JuchemCassingGreiner2004,LindnerMuller2006}. 
Random rescattering of pions may still play some role in the space-time evolution of 
the chiral condensate and the freeze-out process.   
In this regard, it is interesting to see how our "collisionless" pion mode is converted to the 
Son-Stephanov pion mode\cite{SonStephanov2002} which is an isospin counterpart of 
the hydrodynamic spin wave\cite{HalperinHohenberg1969}.   
Our present formalism emphasizes the role of the mesonic mean field in the final state interaction.  
We like to apply our formalism to the kinetic freeze-out dynamics in consideration for the effect of the 
expansion of the system and solve the space-time evolution of the condensate assuming boost 
invariance of the system. 
It is interesting to see whether the Vlasov term might generate strong flow effect due to the gradient 
of pionic mean field.  This question may be studied by the method developed in this work.

One of the important features of our formalism is that it can describe the chiral phase transition 
both in equilibrium and far out of equilibrium, including freeze-out process into 
free streaming particles, in a unified fashion. 
This is desirable for extracting observable consequence of the chiral phase transition 
to confront with experiments of high energy heavy ion collisions which are very complex 
non-equilibrium phenomenon. 

\vspace{1cm}
\centerline {\large \textbf{Acknowledgments}}

We thank H. Fujii, T. Hatsuda, T. Hirano, K. Itakura, Y. Kato, T. Kita,  O. Morimatsu, T. Nikuni 
and Y. Tsue for useful comments and their interests in this work.   
In particular, we are grateful to Yasuhiko Tsue for informing us of their related new results before 
publication.  

\newpage
\appendix
%\section{Supplementary explanations for one-component real scalar field theory}
\section{General relations of the Wigner functions}
General relations between the four components of the Wigner functions which follow 
immediately from the definitions 
(\ref{Fij}), (\ref{Fbarij}), (\ref{Gij}) and (\ref{Gbarij}) are:
\begin{eqnarray}
F_{ij}^{\ast}(\v{p},\v{k},t) &=& F_{ji} (\v{p},-\v{k},t) , 
\label{Fijast}\\  
\bar{F}_{ij}^{\ast}(\v{p},\v{k},t) &=& \bar{F}_{ji} (\v{p},-\v{k},t) , 
\label{barFijast} \\  
\bar{G}_{ij}(\v{p},\v{k},t) &=& G_{ji}^{\ast} (\v{p},-\v{k},t) , 
\label{barGijast}
\end{eqnarray}
where the asterisk (*) stands for the complex conjugate. 
From the equal-time commutation relations of the quantum fields, (\ref{com1}) and 
(\ref{com2}), we obtain the additional relations between the Wigner functions
\begin{eqnarray}
\bar{F}_{ij}(\v{p},\v{k},t) &=& F_{ji}(-\v{p},\v{k},t) + \delta_{ij}\delta_{\v{k}, 0} \label{FijFbarij} 
\label{AFij-barFij}\\
G_{ij}(\v{p},\v{k},t) &=& G_{ji}(-\v{p},\v{k},t) 
\label{AGij}\\  
\bar{G}_{ij}(\v{p},\v{k},t) &=& \bar{G}_{ji}(-\v{p},\v{k},t) 
\label{AGbarij}
\end{eqnarray}

We define the Fourier transforms of the Wigner functions as
\begin{eqnarray}
f_{ij}\prt &=& \sum_{\v{k}} e^{-i\v{k} \cdot \v{r}} F_{ij}(\v{p},\v{k},t),\\ 
 \bar{f}_{ij} \prt & = & \sum_{\v{k}} e^{-i\v{k} \cdot \v{r}} \bar{F}_{ij}(\v{p},\v{k},t),  \\  
g_{ij}\prt &=& \sum_{\v{k}} e^{-i\v{k} \cdot \v{r}} G_{ij}(\v{p},\v{k},t),  \\
 \bar{g}_{ij}\prt & = & \sum_{\v{k}} e^{-i\v{k} \cdot \v{r}} \bar{G}_{ij}(\v{p},\v{k},t), 
\end{eqnarray}

Then (\ref{Fijast}) and (\ref{barFijast}) imply that the diagonal components 
$f_{ii}(\v{p},\v{r},t)$ and $\bar{f}_{ii}(\v{p},\v{r},t)$ are both real functions and 
(\ref{AFij-barFij}) imply that 
\begin{eqnarray}
\bar{f}_{ij}(\v{p},\v{r},t) = f_{ji}(-\v{p},\v{r},t) + \delta_{ij}
\end{eqnarray}
while (\ref{barGijast}) implies that $g_{ij}(\v{p},\v{r},t)$ and $\bar{g}_{ji}(\v{p},\v{r},t)$ are 
complex conjugate to each other:
\begin{eqnarray}
g_{ij}^{\ast}(\v{p},\v{r},t) = \bar{g}_{ji} (\v{p},\v{r},t)
\end{eqnarray}

%Appendix B
\section{Equation of motion of the Wigner functions}

The equation of motion of each component of the Wigner functions is given below:

\begin{eqnarray}
%\hspace{-1cm} 
i \frac{\partial}{\partial t} F_{ij}(\v{p},\v{k},t)
&= &-(\omega_{i,\v{p}+\frac{\v{k}}{2}} - \omega_{j,\v{p}-\frac{\v{k}}{2}})  F_{ij}(\v{p},\v{k},t) 
\nonumber \\ 
&&\hspace{-1cm}  + \sum_{l,\v{q}} 
 \biggl\{ 
 - \Delta \Pi_{il,\v{q}}
 \frac{F_{lj}(\v{p}+\frac{\v{q}}{2} \, , \, \v{k}+\v{q} \, , t) 
 + G_{lj}(\v{p}+\frac{\v{q}}{2} \, , \, \v{k}+\v{q} \, , t)}
{2\sqrt{\omega_{i,\, \v{p}+\frac{\v{k}}{2}}\omega_{l, \, \v{p}+\frac{\v{k}}{2}+\v{q}}}} \nonumber  \\  
&& %\hspace{cm}  
 + \frac{F_{il}(\v{p}-\frac{\v{q}}{2} \, , \, \v{k}+\v{q} \, , t) 
 + \bar{G}_{il}(\v{p}-\frac{\v{q}}{2} \, , \, \v{k}+\v{q} \, , t)}
{2\sqrt{\omega_{l,\, \v{p}-\frac{\v{k}}{2}}\omega_{j, \, \v{p}-\frac{\v{k}}{2}-\v{q}}}}\Delta \Pi_{lj,\v{q}}
 \biggr\}
 \nonumber \\
 \\
i \frac{\partial}{\partial t} G_{ij}(\v{p},\v{k},t) 
&=& (\omega_{i,\v{p}+\frac{\v{k}}{2}} + \omega_{j,\v{p}-\frac{\v{k}}{2}}) 
 G_{ij}(\v{p},\v{k},t) \nonumber \\
&&\hspace{-1cm} +\sum_{l,\v{q}}  
 \biggl\{ 
 \Delta \Pi_{il,\v{q}}
 \frac{
 \,  \left( G_{lj}(\v{p}+\frac{\v{k}}{2}  ,  \v{k}+\v{q}  , t) 
 + F_{lj}(\v{p}+\frac{\v{k}}{2}  ,  \v{k}+\v{q}  , t) \right)}
{\sqrt{2\omega_{i,\, \v{p}+\frac{\v{k}}{2}}}
\sqrt{2\omega_{l, \, \v{p}+\frac{\v{k}}{2}+\v{q}}}} 
  \nonumber  \\
&&%\hspace{-2cm} +
 + \frac{G_{il}(\v{p}-\frac{\v{k}}{2} \, , \, \v{k}+\v{q} \, , t) 
 + \bar{F}_{il}(\v{p}-\frac{\v{q}}{2} \, , \, \v{k}+\v{q} \, , t)}
{\sqrt{2\omega_{l,\, \v{p}-\frac{\v{k}}{2}}}\sqrt{2\omega_{j, \, \v{p}-\frac{\v{k}}{2}-\v{q}}}}\Delta \Pi_{lj,\v{q}}
 \biggr\}
 \nonumber \\
 \\ 
i \frac{\partial}{\partial t} \bar{G}_{ij}(\v{p},\v{k},t) 
 &=& -(\omega_{i,\v{p}+\frac{\v{k}}{2}} + \omega_{j,\v{p}-\frac{\v{k}}{2}}) 
 \bar{G}_{ij}(\v{p},\v{k},t) \nonumber \\
&&\hspace{-1cm} +\sum_{l,\v{q}} 
 \biggl\{ 
 -\Delta \Pi_{il,\v{q}}
 \frac{
 \,  \left( \bar{G}_{lj}(\v{p}+\frac{\v{q}}{2}  ,  \v{k}+\v{q}  , t) 
 + \bar{F}_{lj}(\v{p}+\frac{\v{q}}{2}  ,  \v{k}+\v{q}  , t) \right)}
{\sqrt{2\omega_{i,\, \v{p}+\frac{\v{k}}{2}}}
\sqrt{2\omega_{l, \, \v{p}+\frac{\v{k}}{2}+\v{q}}}} 
\nonumber \\
&&%\hspace{-2cm} -
 - \frac{\bar{G}_{il}(\v{p}-\v{q}/2 \, , \, \v{k}+\v{q} \, , t) 
 + F_{il}(\v{p}-\v{q}/2 \, , \, \v{k}+\v{q} \, , t)}
{\sqrt{2\omega_{l,\, \v{p}-\v{k}/2}}\sqrt{2\omega_{j, \, \v{p}-\v{k}/2-\v{q}}}}\Delta \Pi_{lj,\v{q}}
 \biggr\}
 \nonumber \\
 \\
i \frac{\partial}{\partial t} \bar{F}_{ij}(\v{p},\v{k},t) 
 &=& (\omega_{i,\v{p}+\frac{\v{k}}{2}} - \omega_{j,\v{p}-\frac{\v{k}}{2}}) \bar{F}_{ij}(\v{p},\v{k},t) 
 \nonumber \\
\,  &&\hspace{-1cm} + \sum_{j',\v{q}} 
 \biggl\{ 
 \Delta \Pi_{il,\v{q}}
 \frac{
 \,  \left( \bar{F}_{lj}(\v{p}+\frac{\v{q}}{2}  ,  \v{k}+\v{q}  , t) 
 + \bar{G}_{lj}(\v{p}+\frac{\v{q}}{2}  ,  \v{k}+\v{q}  , t) \right)}
{\sqrt{2\omega_{i,\, \v{p}+\frac{\v{k}}{2}}}
\sqrt{2\omega_{l, \, \v{p}+\frac{\v{k}}{2}+\v{q}}}} 
  \nonumber  \\
&&%\hspace{-2cm}-
 -\frac{\bar{F}_{il}(\v{p}-\v{q}/2 \, , \, \v{k}+\v{q} \, , t) 
 + G_{il}(\v{p}-\v{q}/2 \, , \, \v{k}+\v{q} \, , t)}
{\sqrt{2\omega_{l,\, \v{p}-\v{k}/2}}\sqrt{2\omega_{j, \, \v{p}-\v{k}/2-\v{q}}}}\Delta \Pi_{lj,\v{q}}
 \biggr\}
 \nonumber \\
\end{eqnarray}

\section{Kinetic equations of $O(N)$ model in the long wavelength approximation}
Here we present the explicit form of the equations of motion of the Wigner functions 
in the long wavelength approximation: 
\begin{eqnarray} 
&& i \dot{f}_{ij}(\v{p},\v{r},t)= - (\omega_{i,\v{p}} - \omega_{j,\v{p}}) f_{ij}(\v{p},\v{r},t) - \frac{1}{2} \left( \frac{1}{\omega_{i,\v{p}}} + \frac{1}{\omega_{j,\v{p}}} \right) \v{p} \cdot (i \nabla_{\v{r}} f_{ij}(\v{p},\v{r},t))  \nonumber \\
&&  + \sum_{j'} 
\Biggl[
- \biggl\{
U_{ij'}(\v{r},t) f_{j'j}(\v{p},\v{r},t) 
- \frac{1}{4} \left( \frac{1}{\omega^2_{i,\v{p}}} - \frac{1}{\omega^2_{j',\v{p}}} \right) \v{p} \cdot (i \nabla _{\v{r}} U_{ij'}(\v{r},t) ) f_{j'j}(\v{p},\v{r},t) \nonumber \\
&& \hspace{2cm}  
- \frac{1}{4} \left( \frac{1}{\omega^2_{i,\v{p}}} + \frac{1}{\omega^2_{j',\v{p}}} \right) U_{ij'}(\v{r},t) \v{p} \cdot (i \nabla _{\v{r}}  f_{j'j}(\v{p},\v{r},t) ) \nonumber \\
&& \hspace{2cm}
+ \frac{1}{2} ( - i \nabla_{\v{r}} U_{ij'}(\v{r},t) \cdot \nabla_{\v{p}} f_{j'j}(\v{p},\v{r},t) ) 
\biggr\} \nonumber \\
&& - \biggl\{
U_{ij'}(\v{r},t) g_{j'j}(\v{p},\v{r},t) 
- \frac{1}{4} \left( \frac{1}{\omega^2_{i,\v{p}}} - \frac{1}{\omega^2_{j',\v{p}}} \right) \v{p} \cdot (i \nabla _{\v{r}} U_{ij'}(\v{r},t) ) g_{j'j}(\v{p},\v{r},t) \nonumber \\
 && \hspace{2cm}  
- \frac{1}{4} \left( \frac{1}{\omega^2_{i,\v{p}}} + \frac{1}{\omega^2_{j',\v{p}}} \right) U_{ij'}(\v{r},t) \v{p} \cdot (i \nabla _{\v{r}}  g_{j'j}(\v{p},\v{r},t) ) \nonumber \\
&& \hspace{2cm}
+ \frac{1}{2} ( - i \nabla_{\v{r}} U_{ij'}(\v{r},t) \cdot \nabla_{\v{p}} g_{j'j}(\v{p},\v{r},t) ) \biggr\}
 \nonumber \\
&& + \biggl\{
U_{j'j}(\v{r},t) f_{ij'}(\v{p},\v{r},t) 
+ \frac{1}{4} \left( \frac{1}{\omega^2_{i,\v{p}}} - \frac{1}{\omega^2_{j',\v{p}}} \right) \v{p} \cdot (i \nabla _{\v{r}} U_{j'j}(\v{r},t) ) f_{ij'}(\v{p},\v{r},t) \nonumber \\
&& \hspace{2cm}  
+ \frac{1}{4} \left( \frac{1}{\omega^2_{i,\v{p}}} + \frac{1}{\omega^2_{j',\v{p}}} \right) U_{j'j}(\v{r},t) \v{p} \cdot (i \nabla _{\v{r}}  f_{ij'}(\v{p},\v{r},t) ) \nonumber \\
&& \hspace{2cm}
- \frac{1}{2} ( - i \nabla_{\v{r}} U_{j'j}(\v{r},t) \cdot \nabla_{\v{p}} f_{ij'}(\v{p},\v{r},t) ) 
\biggr\} \nonumber \\
&& +  \biggl\{
U_{j'j}(\v{r},t) \bar{g}_{ij'}(\v{p},\v{r},t) 
+ \frac{1}{4} \left( \frac{1}{\omega^2_{i,\v{p}}} - \frac{1}{\omega^2_{j',\v{p}}} \right) \v{p} \cdot (i \nabla _{\v{r}} U_{j'j}(\v{r},t) ) \bar{g}_{ij'}(\v{p},\v{r},t) \nonumber \\
&& \hspace{2cm}  
+ \frac{1}{4} \left( \frac{1}{\omega^2_{i,\v{p}}} + \frac{1}{\omega^2_{j',\v{p}}} \right) U_{j'j}(\v{r},t) \v{p} \cdot (i \nabla _{\v{r}}  \bar{g}_{ij'}(\v{p},\v{r},t) ) \nonumber \\
&& \hspace{2cm}
- \frac{1}{2} ( - i \nabla_{\v{r}} U_{j'j}(\v{r},t) \cdot \nabla_{\v{p}} \bar{g}_{ij'}(\v{p},\v{r},t) ) 
\biggr\} \Biggr]
\nonumber \\
\end{eqnarray}
 
\begin{eqnarray} 
&& i \dot{\bar{f}}_{ij}(\v{p},\v{r},t)= + (\omega_{i,\v{p}} - \omega_{j,\v{p}}) \bar{f}_{ij}(\v{p},\v{r},t) + \frac{1}{2} \left( \frac{1}{\omega_{i,\v{p}}} + \frac{1}{\omega_{j,\v{p}}} \right) \v{p} \cdot (i \nabla_{\v{r}} \bar{f}_{ij}(\v{p},\v{r},t))  \nonumber \\
&&  - \sum_{j'} 
\Biggl[
- \biggl\{
U_{ij'}(\v{r},t) \bar{f}_{j'j}(\v{p},\v{r},t) 
- \frac{1}{4} \left( \frac{1}{\omega^2_{i,\v{p}}} - \frac{1}{\omega^2_{j',\v{p}}} \right) \v{p} \cdot (i \nabla _{\v{r}} U_{ij'}(\v{r},t) ) \bar{f}_{j'j}(\v{p},\v{r},t) \nonumber \\
&& \hspace{2cm}  
- \frac{1}{4} \left( \frac{1}{\omega^2_{i,\v{p}}} + \frac{1}{\omega^2_{j',\v{p}}} \right) U_{ij'}(\v{r},t) \v{p} \cdot (i \nabla _{\v{r}}  \bar{f}_{j'j}(\v{p},\v{r},t) ) \nonumber \\
&& \hspace{2cm}
+ \frac{1}{2} ( - i \nabla_{\v{r}} U_{ij'}(\v{r},t) \cdot \nabla_{\v{p}} \bar{f}_{j'j}(\v{p},\v{r},t) ) 
\biggr\} \nonumber \\
&& - \biggl\{
U_{ij'}(\v{r},t) \bar{g}_{j'j}(\v{p},\v{r},t) 
- \frac{1}{4} \left( \frac{1}{\omega^2_{i,\v{p}}} - \frac{1}{\omega^2_{j',\v{p}}} \right) \v{p} \cdot (i \nabla _{\v{r}} U_{ij'}(\v{r},t) ) \bar{g}_{j'j}(\v{p},\v{r},t) \nonumber \\
 && \hspace{2cm}  
- \frac{1}{4} \left( \frac{1}{\omega^2_{i,\v{p}}} + \frac{1}{\omega^2_{j',\v{p}}} \right) U_{ij'}(\v{r},t) \v{p} \cdot (i \nabla _{\v{r}}  \bar{g}_{j'j}(\v{p},\v{r},t) ) \nonumber \\
&& \hspace{2cm}
+ \frac{1}{2} ( - i \nabla_{\v{r}} U_{ij'}(\v{r},t) \cdot \nabla_{\v{p}} \bar{g}_{j'j}(\v{p},\v{r},t) ) \biggr\}
 \nonumber \\
&& + \biggl\{
U_{j'j}(\v{r},t) \bar{f}_{ij'}(\v{p},\v{r},t) 
+ \frac{1}{4} \left( \frac{1}{\omega^2_{i,\v{p}}} - \frac{1}{\omega^2_{j',\v{p}}} \right) \v{p} \cdot (i \nabla _{\v{r}} U_{j'j}(\v{r},t) ) \bar{f}_{ij'}(\v{p},\v{r},t) \nonumber \\
&& \hspace{2cm}  
+ \frac{1}{4} \left( \frac{1}{\omega^2_{i,\v{p}}} + \frac{1}{\omega^2_{j',\v{p}}} \right) U_{j'j}(\v{r},t) \v{p} \cdot (i \nabla _{\v{r}}  \bar{f}_{ij'}(\v{p},\v{r},t) ) \nonumber \\
&& \hspace{2cm}
- \frac{1}{2} ( - i \nabla_{\v{r}} U_{j'j}(\v{r},t) \cdot \nabla_{\v{p}} \bar{f}_{ij'}(\v{p},\v{r},t) ) 
\biggr\} \nonumber \\
&& + \biggl\{
U_{j'j}(\v{r},t) g_{ij'}(\v{p},\v{r},t) 
+ \frac{1}{4} \left( \frac{1}{\omega^2_{i,\v{p}}} - \frac{1}{\omega^2_{j',\v{p}}} \right) \v{p} \cdot (i \nabla _{\v{r}} U_{j'j}(\v{r},t) ) g_{ij'}(\v{p},\v{r},t) \nonumber \\
&& \hspace{2cm}  
+ \frac{1}{4} \left( \frac{1}{\omega^2_{i,\v{p}}} + \frac{1}{\omega^2_{j',\v{p}}} \right) U_{j'j}(\v{r},t) \v{p} \cdot (i \nabla _{\v{r}}  g_{ij'}(\v{p},\v{r},t) ) \nonumber \\
&& \hspace{2cm}
- \frac{1}{2} ( - i \nabla_{\v{r}} U_{j'j}(\v{r},t) \cdot \nabla_{\v{p}} g_{ij'}(\v{p},\v{r},t) ) 
\biggr\} \Biggr]
\end{eqnarray}

\begin{eqnarray} 
&& i \dot{g}_{ij}(\v{p},\v{r},t)= - (\omega_{i,\v{p}} + \omega_{j,\v{p}}) f_{ij}(\v{p},\v{r},t) - \frac{1}{2} \left( \frac{1}{\omega_{i,\v{p}}} - \frac{1}{\omega_{j,\v{p}}} \right) \v{p} \cdot (i \nabla_{\v{r}} g_{ij}(\v{p},\v{r},t))  \nonumber \\
&&  + \sum_{j'} 
\Biggl[
 \biggl\{
U_{ij'}(\v{r},t) g_{j'j}(\v{p},\v{r},t) 
- \frac{1}{4} \left( \frac{1}{\omega^2_{i,\v{p}}} - \frac{1}{\omega^2_{j',\v{p}}} \right) \v{p} \cdot (i \nabla _{\v{r}} U_{ij'}(\v{r},t) ) g_{j'j}(\v{p},\v{r},t) \nonumber \\
&& \hspace{2cm}  
- \frac{1}{4} \left( \frac{1}{\omega^2_{i,\v{p}}} + \frac{1}{\omega^2_{j',\v{p}}} \right) U_{ij'}(\v{r},t) \v{p} \cdot (i \nabla _{\v{r}}  g_{j'j}(\v{p},\v{r},t) ) \nonumber \\
&& \hspace{2cm}
+ \frac{1}{2} ( - i \nabla_{\v{r}} U_{ij'}(\v{r},t) \cdot \nabla_{\v{p}} g_{j'j}(\v{p},\v{r},t) ) 
\biggr\} \nonumber \\
&& + \biggl\{
U_{ij'}(\v{r},t) f_{j'j}(\v{p},\v{r},t) 
- \frac{1}{4} \left( \frac{1}{\omega^2_{i,\v{p}}} - \frac{1}{\omega^2_{j',\v{p}}} \right) \v{p} \cdot (i \nabla _{\v{r}} U_{ij'}(\v{r},t) ) f_{j'j}(\v{p},\v{r},t) \nonumber \\
 && \hspace{2cm}  
- \frac{1}{4} \left( \frac{1}{\omega^2_{i,\v{p}}} + \frac{1}{\omega^2_{j',\v{p}}} \right) U_{ij'}(\v{r},t) \v{p} \cdot (i \nabla _{\v{r}}  f_{j'j}(\v{p},\v{r},t) ) \nonumber \\
&& \hspace{2cm}
+ \frac{1}{2} ( - i \nabla_{\v{r}} U_{ij'}(\v{r},t) \cdot \nabla_{\v{p}} f_{j'j}(\v{p},\v{r},t) ) \biggr\}
 \nonumber \\
&& + \biggl\{
U_{j'j}(\v{r},t) g_{ij'}(\v{p},\v{r},t) 
+ \frac{1}{4} \left( \frac{1}{\omega^2_{i,\v{p}}} - \frac{1}{\omega^2_{j',\v{p}}} \right) \v{p} \cdot (i \nabla _{\v{r}} U_{j'j}(\v{r},t) ) g_{ij'}(\v{p},\v{r},t) \nonumber \\
&& \hspace{2cm}  
+ \frac{1}{4} \left( \frac{1}{\omega^2_{i,\v{p}}} + \frac{1}{\omega^2_{j',\v{p}}} \right) U_{j'j}(\v{r},t) \v{p} \cdot (i \nabla _{\v{r}}  g_{ij'}(\v{p},\v{r},t) ) \nonumber \\
&& \hspace{2cm}
- \frac{1}{2} ( - i \nabla_{\v{r}} U_{j'j}(\v{r},t) \cdot \nabla_{\v{p}} g_{ij'}(\v{p},\v{r},t) ) 
\biggr\} \nonumber \\
&& + \biggl\{
U_{j'j}(\v{r},t) \bar{f}_{ij'}(\v{p},\v{r},t) 
+ \frac{1}{4} \left( \frac{1}{\omega^2_{i,\v{p}}} - \frac{1}{\omega^2_{j',\v{p}}} \right) \v{p} \cdot (i \nabla _{\v{r}} U_{j'j}(\v{r},t) ) \bar{f}_{ij'}(\v{p},\v{r},t) \nonumber \\
&& \hspace{2cm}  
+ \frac{1}{4} \left( \frac{1}{\omega^2_{i,\v{p}}} + \frac{1}{\omega^2_{j',\v{p}}} \right) U_{j'j}(\v{r},t) \v{p} \cdot (i \nabla _{\v{r}}  \bar{f}_{ij'}(\v{p},\v{r},t) ) \nonumber \\
&& \hspace{2cm}
- \frac{1}{2} ( - i \nabla_{\v{r}} U_{j'j}(\v{r},t) \cdot \nabla_{\v{p}} \bar{f}_{ij'}(\v{p},\v{r},t) ) 
\biggr\} \Biggr]
\end{eqnarray}% \g_{ij}ã?Ê?ã?Ê?Vlasovã?Ê?ã?Ê?ã?Ê?ã?ÊÝã?Ê?ã?Ê?
\begin{eqnarray} % \bar{g} ij ã?Ê?ã?Ê?ã?Ê?ã?Ê?ã?Ê?ã?ÊÝã?Ê?ã?Ê?
&& i \dot{\bar{g}}_{ij}(\v{p},\v{r},t)= + (\omega_{i,\v{p}} + \omega_{j,\v{p}}) \bar{f}_{ij}(\v{p},\v{r},t) + \frac{1}{2} \left( \frac{1}{\omega_{i,\v{p}}} - \frac{1}{\omega_{j,\v{p}}} \right) \v{p} \cdot (i \nabla_{\v{r}} \bar{g}_{ij}(\v{p},\v{r},t))  \nonumber \\
&&  - \sum_{j'} 
\Biggl[
 \biggl\{
U_{ij'}(\v{r},t) \bar{g}_{j'j}(\v{p},\v{r},t) 
- \frac{1}{4} \left( \frac{1}{\omega^2_{i,\v{p}}} - \frac{1}{\omega^2_{j',\v{p}}} \right) \v{p} \cdot (i \nabla _{\v{r}} U_{ij'}(\v{r},t) ) \bar{g}_{j'j}(\v{p},\v{r},t) \nonumber \\
&& \hspace{2cm}  
- \frac{1}{4} \left( \frac{1}{\omega^2_{i,\v{p}}} + \frac{1}{\omega^2_{j',\v{p}}} \right) U_{ij'}(\v{r},t) \v{p} \cdot (i \nabla _{\v{r}}  \bar{g}_{j'j}(\v{p},\v{r},t) ) \nonumber \\
&& \hspace{2cm}
+ \frac{1}{2} ( - i \nabla_{\v{r}} U_{ij'}(\v{r},t) \cdot \nabla_{\v{p}} \bar{g}_{j'j}(\v{p},\v{r},t) ) 
\biggr\} \nonumber \\
&& + \biggl\{
U_{ij'}(\v{r},t) \bar{f}_{j'j}(\v{p},\v{r},t) 
- \frac{1}{4} \left( \frac{1}{\omega^2_{i,\v{p}}} - \frac{1}{\omega^2_{j',\v{p}}} \right) \v{p} \cdot (i \nabla _{\v{r}} U_{ij'}(\v{r},t) ) \bar{f}_{j'j}(\v{p},\v{r},t) \nonumber \\
 && \hspace{2cm}  
- \frac{1}{4} \left( \frac{1}{\omega^2_{i,\v{p}}} + \frac{1}{\omega^2_{j',\v{p}}} \right) U_{ij'}(\v{r},t) \v{p} \cdot (i \nabla _{\v{r}}  \bar{f}_{j'j}(\v{p},\v{r},t) ) \nonumber \\
&& \hspace{2cm}
+ \frac{1}{2} ( - i \nabla_{\v{r}} U_{ij'}(\v{r},t) \cdot \nabla_{\v{p}} \bar{f}_{j'j}(\v{p},\v{r},t) ) \biggr\}
 \nonumber \\
&& + \biggl\{
U_{j'j}(\v{r},t) \bar{g}_{ij'}(\v{p},\v{r},t) 
+ \frac{1}{4} \left( \frac{1}{\omega^2_{i,\v{p}}} - \frac{1}{\omega^2_{j',\v{p}}} \right) \v{p} \cdot (i \nabla _{\v{r}} U_{j'j}(\v{r},t) ) \bar{g}_{ij'}(\v{p},\v{r},t) \nonumber \\
&& \hspace{2cm}  
+ \frac{1}{4} \left( \frac{1}{\omega^2_{i,\v{p}}} + \frac{1}{\omega^2_{j',\v{p}}} \right) U_{j'j}(\v{r},t) \v{p} \cdot (i \nabla _{\v{r}}  \bar{g}_{ij'}(\v{p},\v{r},t) ) \nonumber \\
&& \hspace{2cm}
- \frac{1}{2} ( - i \nabla_{\v{r}} U_{j'j}(\v{r},t) \cdot \nabla_{\v{p}} \bar{g}_{ij'}(\v{p},\v{r},t) ) 
\biggr\} \nonumber \\
&& + \biggl\{
U_{j'j}(\v{r},t) f_{ij'}(\v{p},\v{r},t) 
+ \frac{1}{4} \left( \frac{1}{\omega^2_{i,\v{p}}} - \frac{1}{\omega^2_{j',\v{p}}} \right) \v{p} \cdot (i \nabla _{\v{r}} U_{j'j}(\v{r},t) ) f_{ij'}(\v{p},\v{r},t) \nonumber \\
&& \hspace{2cm}  
+ \frac{1}{4} \left( \frac{1}{\omega^2_{i,\v{p}}} + \frac{1}{\omega^2_{j',\v{p}}} \right) U_{j'j}(\v{r},t) \v{p} \cdot (i \nabla _{\v{r}}  f_{ij'}(\v{p},\v{r},t) ) \nonumber \\
&& \hspace{2cm}
- \frac{1}{2} ( - i \nabla_{\v{r}} U_{j'j}(\v{r},t) \cdot \nabla_{\v{p}} f_{ij'}(\v{p},\v{r},t) ) 
\biggr\} \Biggr]
\end{eqnarray}

\section{Linearized Vlasov equations of $O(N)$ model}
The complete set of the linearized Vlasov equations discussed in section 3 is listed below: 
\begin{eqnarray}
i \delta \dot{f}_{ij}(\v{p},\v{r},t)
  &=& - (\omega_{i,\p} - \omega_{j,\p}) \delta f_{ij}(\v{p},\v{r},t)
  - \frac{i}{2}\big( \frac{1}{\omega_{i,\p}} + \frac{1}{\omega_{j,\p}} \big)
  \v{p} \cdot \nabla_{\bf r}  \delta f_{ij} \nonumber \\
 & & \hspace{0.3cm} +
( \foi -  \foj) \left[ \delta U_{ij}(\v{r},t) 
 - \frac{i}{4}\big( \frac{1}{\omega_{i,\p}^2} - \frac{1}{\omega_{j,\p}^2} \big)
   \v{p} \cdot \nabla_{\bf r}  \delta U_{ij} \right]
      \nonumber \\  & & \hspace{0.6cm} + 
  \frac{i}{2} \nabla_{\v{p}} ( \foi + \foj) \cdot \nabla_{\bf r}  \delta U_{ij} 
  \nonumber \\
  \\
i \delta \dot{\bar{f}}_{ij}(\v{p},\v{r},t)
  &=& - (\omega_{i,\p} - \omega_{j,\p}) \delta \bar{f}_{ij}(\v{p},\v{r},t)
  - \frac{i}{2}\big( \frac{1}{\omega_{i,\p}} + \frac{1}{\omega_{j,\p}} \big)
  \v{p} \cdot  \nabla_{\bf r} \delta \bar{f}_{ij} \nonumber \\
& & \hspace{0.3cm} -  ( \foi -  \foj)
\left[ \delta U_{ij}(\v{r},t) 
 - \frac{i}{4}\big( \frac{1}{\omega_{i,\p}^2} - \frac{1}{\omega_{j,\p}^2} \big)
   \v{p} \cdot \nabla_{\bf r} \delta U_{ij} \right]
   \nonumber \\ & & \hspace{0.6cm}
 - \frac{i}{2} \nabla_{\v{p}} ( \foi +  \foj) \cdot \nabla_{\bf r} \delta U_{ij} 
 \nonumber \\
 \\
 \label{deltabarfij}
i \delta \dot{g}_{ij}(\v{p},\v{r},t)
  &=&  (\omega_{i,\p} + \omega_{j,\p}) \delta g_{ij}(\v{p},\v{r},t)
  + \frac{i}{2}\big( \frac{1}{\omega_{i,\p}} - \frac{1}{\omega_{j,\p}} \big)
  \v{p} \cdot \nabla_{\bf r}  \delta g_{ij} \nonumber \\
& & \hspace{0.3cm} + ( \foi +  \foj)
\left[ \delta U_{ij}(\v{r},t) 
 - \frac{i}{4}\big( \frac{1}{\omega_{i,\p}^2} - \frac{1}{\omega_{j,\p}^2} \big)
   \v{p} \cdot \nabla_{\bf r}  \delta U_{ij}  \right]
      \nonumber \\ & & \hspace{0.6cm}
 + \frac{i}{2}  \nabla_{\v{p}} ( \foi -   \foj) \cdot  \nabla_{\bf r}  \delta U_{ij} 
 \nonumber \\
 \\
i \delta \dot{\bar{g}}_{ij}(\v{p},\v{r},t)
  &=&  - (\omega_{i,\p} + \omega_{j,\p}) \delta \bar{g}_{ij}(\v{p},\v{r},t)
  - \frac{i}{2}\big( \frac{1}{\omega_{i,\p}} - \frac{1}{\omega_{j,\p}} \big)
  \v{p} \cdot  \nabla_{\bf r}  \delta \bar{g}_{ij} \nonumber \\
& &  \hspace{0.3cm} +  ( \foi +   \foj)
\left[ \delta U_{ij}(\v{r},t)
 + \frac{i}{4}\big( \frac{1}{\omega_{i,\p}^2} - \frac{1}{\omega_{j,\p}^2} \big)
   \v{p} \cdot  \nabla_{\bf r}  \delta U_{ij}  \right]
    \nonumber \\ & & \hspace{0.6cm}
 - \frac{i}{2} i  \nabla_{\v{p}} ( \foi -   \foj) \cdot \nabla_{\bf r}  \delta U_{ij} 
\nonumber \\
 \label{deltabargij}
\end{eqnarray}

The Fourier transform of  these equations yields:
\begin{eqnarray}
 & & \hspace{-2cm}
  \Bigl\{
\omega + \omega_{i} - \omega_{j} 
- \frac{\v{p} \cdot \v{k}}{2}\big( \frac{1}{\omega_{i}} + \frac{1}{\omega_{j}} \big) 
\Bigr\}
 \delta f_{ij}(\v{p},\v{k},\omega) \nonumber \\
& &\hspace{-1.6cm}
= 
\biggl[
\Bigl\{  1 + \frac{1}{4}\big( \frac{1}{\omega_{i}^2} - \frac{1}{\omega_{j}^2} \big)  
\v{p} \cdot \v{k} \Bigr\}
( f_{\rm eq.} (i)  - f_{\rm eq.} (j) )
 - \frac{1}{2} \v{k} \cdot \nabla_{\v{p}} ( f_{\rm eq.} (i)  + f_{\rm eq.} (j) )
\biggr]  \delta U_{ij} (\v{k},\omega)  \nonumber \\
\label{deltaWij1}
\\ 
& & \hspace{-2cm}
     \Bigl\{  
\omega - \omega_{i} + \omega_{j}
+ \frac{\v{p} \cdot \v{k}}{2}\big( \frac{1}{\omega_{i}} + \frac{1}{\omega_{j}} \big) 
\Bigr\}
 \delta \bar{f}_{ij}(\v{p},\v{k},\omega) \nonumber \\
& &\hspace{-1.6cm}
=  
\biggl[ - 
\Bigl\{  1 + \frac{1}{4}\big( \frac{1}{\omega_{i}^2} - \frac{1}{\omega_{j}^2} \big)  
\v{p} \cdot \v{k} \Bigr\}
( f_{\rm eq.} (i)  - f_{\rm eq.} (j) )
 + \frac{1}{2} \v{k} \cdot \nabla_{\v{p}} ( f_{\rm eq.} (i)  + f_{\rm eq.} (j) )
\biggr]   \delta U_{ij} (\v{k},\omega) \nonumber \\ \label{deltaWij2}
\\
& & \hspace{-2cm}
  \Bigl\{  
\omega - \omega_{i} - \omega_{j} 
+ \frac{\v{p} \cdot \v{k}}{2}\big( \frac{1}{\omega_{i}} - \frac{1}{\omega_{j}} \big) 
\Bigr\}
 \delta g_{ij}(\v{p},\v{k},\omega) \nonumber  \\
& &\hspace{-1.6cm}
=  
\biggl[  
\Bigl\{  1 + \frac{1}{4}\big( \frac{1}{\omega_{i}^2} - \frac{1}{\omega_{j}^2} \big)  
\v{p} \cdot \v{k} \Bigr\}
( f_{\rm eq.} (i)  + f_{\rm eq.} (j) )
 - \frac{1}{2} \v{k} \cdot \nabla_{\v{p}} ( f_{\rm eq.} (i)  - f_{\rm eq.} (j) )
\biggr]    \delta U_{ij} (\v{k},\omega) \nonumber \\ \label{deltaWij3}
\\
& & \hspace{-2cm}
  \Bigl\{  
\omega + \omega_{i} + \omega_{j} 
+ \frac{\v{p} \cdot \v{k}}{2}\big( \frac{1}{\omega_{i}} - \frac{1}{\omega_{j}} \big) 
\Bigr\}
 \delta \bar{g}_{ij}(\v{p},\v{k},\omega) \nonumber \\
& &\hspace{-1.6cm}
=  
\biggl[ - 
\Bigl\{  1 + \frac{1}{4}\big( \frac{1}{\omega_{i}^2} - \frac{1}{\omega_{j}^2} \big)  
\v{p} \cdot \v{k} \Bigr\}
( f_{\rm eq.} (i)  + f_{\rm eq.} (j) )
 - \frac{1}{2} \v{k} \cdot \nabla_{\v{p}} ( f_{\rm eq.} (i)  - f_{\rm eq.} (j) )
\biggr]   \delta U_{ij} (\v{k},\omega) \nonumber \\ \label{deltaWij4}
\end{eqnarray}
where we have used slightly abbreviate notations: $\omega_i = \omega_{i,\p} $,
$\omega_j = \omega_{j,\p}$, $f_{\rm eq.} (i) = \foi$ and  $f_{\rm eq.} (j) = \foj$.  

\newpage

\bibliographystyle{unsrt.bst}

\end{document}